\documentclass[preprint]{elsarticle}

\usepackage{mathptmx}
\usepackage{helvet}
\usepackage{courier}
\usepackage{multicol}
\usepackage[bottom]{footmisc}
\usepackage{amssymb}
\usepackage{rotating}

\usepackage{epsfig}
\usepackage{latexsym}
\usepackage{graphicx}
\usepackage{color}
\usepackage{url}
\usepackage{amsfonts}
\usepackage{pgf,pgfarrows,pgfnodes,pgfautomata,pgfheaps,pgfshade}
\usepackage{tikz}
\usetikzlibrary{arrows,automata,positioning}

\usetikzlibrary{arrows,decorations.pathmorphing,backgrounds,positioning,fit,petri}
\usepackage{etex}

\usepackage{stmaryrd}
\usepackage{amsmath}

\usepackage{bussproofs}

\newdefinition{definition}{Definition}
\newdefinition{lemma}{Lemma}
\newdefinition{theorem}{Theorem}
\newdefinition{corollary}{Corollary}
\newdefinition{proposition}{Proposition}
\newdefinition{example}{Example}
\newdefinition{remark}{Remark}
\newdefinition{fact}{Fact}
\newproof{proof}{Proof}

\newcommand{\cons}{\mathcal{C}}

\newcommand{\fine}{{\mbox{ }\nolinebreak\hfill{$\Box$}}}

\newcommand{\len}{\mbox{$len$}}

\newcommand{\deriv}[1]{{\mbox{${\:\stackrel{#1}{\longrightarrow}\:}$}}}

\newcommand{\Deriv}[1]{{\mbox{${\:\stackrel{#1}{\Longrightarrow}\:}$}}}
\newcommand{\derivu}[1]{{\mbox{${\:\stackrel{#1}{\longrightarrow_1}\:}$}}}
\newcommand{\derivd}[1]{{\mbox{${\:\stackrel{#1}{\longrightarrow_2}\:}$}}}

\newcommand{\NDeriv}[1]{{\mbox{${\:\stackrel{#1}{\nRightarrow}\:}$}}}
\newcommand{\nderiv}[1]{\nrightarrow}

\newcommand{\eqdef}{ \doteq }

\newcommand{\bigfrac}[2]{
\renewcommand{\arraystretch}{1.5}
\begin{array}{c}#1\\
\hline
#2
\end{array}}

\newcommand{\nil}{\mbox{\bf 0}}
\newcommand{\1}{\mbox{\bf 1}}

\newcommand{\const}[1]{\mbox{{\it Const}$(#1)$}}

\renewcommand{\mid}{\;\;\big|\;\;}

\newcommand{\encodings}[2]{\ensuremath{\llbracket #2 \rrbracket_{#1}}}

\begin{document}

 \pagestyle{headings}

\begin{frontmatter}
\title{The Algebra of Nondeterministic Finite Automata}

\author{Roberto Gorrieri\corref{cor1}}
\ead{roberto.gorrieri@unibo.it}

\address{Dipartimento di Informatica - Scienza e Ingegneria\\
Universit\`a di Bologna,\\
Mura A. Zamboni 7, 41027 Bologna, Italy}

\cortext[cor1]{Corresponding author}

\begin{abstract}
A process algebra is proposed, whose semantics maps a term to a nondeterministic finite automaton (NFA, for short). We
prove a representability theorem: for each  NFA $N$, there exists a process algebraic term $p$ such that its semantics is an NFA
isomorphic to $N$. Moreover, we provide a concise axiomatization of language equivalence: two NFAs $N_1$ and $N_2$
recognize the same language if and only if the associated terms $p_1$ and $p_2$, respectively, can be equated
by means of a set of axioms, comprising 7 axioms plus 3 conditional axioms, only.
\end{abstract}


\end{frontmatter}

%
\section{Introduction}
%

Nondeterministic finite automata (NFA for short), originally proposed in \cite{RS59}, are a well-known model of sequential 
computation (see, e.g., \cite{HMU01} for an introductory textbook), 
specifically tailored to recognize the class of {\em regular languages}, i.e., those languages that can be
described by means of {\em regular expressions} \cite{Kleene}.

\begin{figure}[t]
\centering
\begin{tikzpicture}
  \tikzset{%
    mythick/.style={%
        line width=.35mm,>=stealth
    }
  }
  \tikzset{%
    mynode/.style={
      circle,
      fill,
      inner sep=2.1pt
    },
    shorten >= 3pt,
    shorten <= 3pt
  }
\def\eodiaglabeldist{0.4mm}
\def\eolabeldist{0mm}
\def\eofigdist{4.5cm}
\def\rrrel{$\cong_r$}
\def\eodist{0.5cm}
\def\eodisty{0.4cm}
\def\eodistw{0.8cm}

\draw [thick,fill=green!25] (-1,-1) circle [radius=1.6cm];
\draw (-1.1,-0.3) node (p0) {NFAs};
\node (p1) [mynode,below =\eodisty of p0, label={[label distance=\eodiaglabeldist]left:$N$}] {};
\node (p2) [mynode,below =\eodist of p1, label={[label distance=\eodiaglabeldist]right:$N'(e_N)$}] {};
 \path (p1) -- node (R3) [inner sep=1pt] {$\sim$} (p2);

\draw [thick,fill=blue!25] (6,-1) circle [radius=1.6cm];
\draw (6,-0.4) node (q0) {Regular Expressions};
\node (q1) [mynode,below =\eodistw of q0, label={[label distance=\eodiaglabeldist]right:$e_N$}] {};

\draw (p1) edge[mythick,->, bend left] node[above] {Kleene's algorithm} (q1);
\draw (q1) edge[mythick,->, bend left] node[below] {McNaughton \& Yamada's algorithm} (p2);

\end{tikzpicture}
\caption{Graphical description of the representability theorem, up to language equivalence}
\label{rep1}
\end{figure}
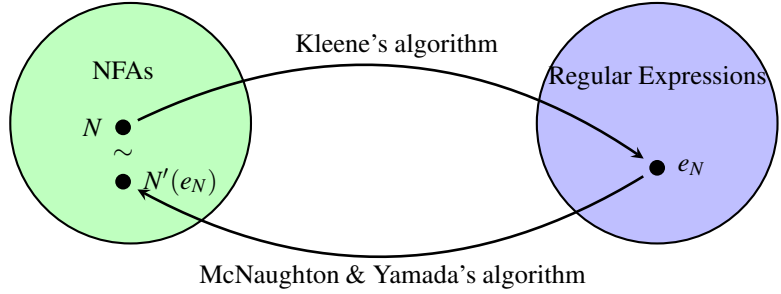

Kleene's algorithm  \cite{Kleene} (see, e.g, \cite{GY04,HMU01} for a more readable and 
accessible description) transforms a given 
nondeterministic finite automaton $N$ into a regular expression $e_N$ that describes the same regular 
language accepted by $N$.  
Conversely, McNaughton and Yamada's algorithm \cite{MNY} converts any regular expression $e$ into an 
equivalent NFA. 
Hence, by a combination of these two results, we get
a {\em Representability Theorem} (see Figure \ref{rep1}) stating that the Kleene algebra of regular expressions truly represents
the class of NFAs, up to language equivalence $\sim$. In fact,
\begin{itemize}
\item[$(i)$] Given an NFA $N$, we construct an equivalent regular expression $e_N$ by Kleene's algorithm: 
so {\em all} NFAs can be represented by regular expressions; and
\item[$(ii)$] from each regular expression $e$ (hence also from $e_N$) we construct an equivalent NFA $N'(e)$ 
(hence, also $N'(e_N)$) by McNaughton and Yamada's algorithm: 
so {\em only} NFAs can be 
represented by regular expressions; and, finally,
\item[$(iii)$] the NFAs $N$ and $N'(e_N)$ are language equivalent, denoted by $N \sim N'(e_N)$.
\end{itemize}

In this paper, we propose an algebra, called SFM1, that also truly represents NFAs, but up to isomorphism.
More precisely, SFM1 is a slight extension to a dialect of finite-state CCS \cite{Mil84,Mil89,GV15}, 
called SFM in \cite{Gor17,Gor20ic}, where an additional constant $\1$ is introduced in order to distinguish 
final states from non-final ones, as introduced in
BPA and related languages (see, e.g., \cite{BBR09} for an extensive investigation of this issue). 
Besides the two constants $\1$ and $\nil$, used to represent a final state and a non-final state, respectively,
SFM1 comprises a disjunction operator (as for regular expressions), denoted by $\_ + \_$ (like the 
choice operator of finite-state CCS), and
a prefixing operator $\alpha.\_$ for each $\alpha \in A \cup \{\epsilon\}$, where $A$ is a 
finite alphabet and $\epsilon$ is the symbol denoting the empty word. The prefixing operator is a weaker form of 
the concatenation operator $\_ \cdot \_$ of regular expressions, because it may concatenate only one single symbol
to a term, like in $a.p$. To compensate the lack of expressivity of this weaker operator, SFM1 replaces the 
iteration operator $\_^*$ (or Kleene star)
of regular expressions with the more powerful recursion operator, implemented by means of process 
constants (as in CCS \cite{Mil89,GV15}), working in the same way as nonterminals are used
in right-linear grammars \cite{HMU01}. For instance, consider a recursive constant $C$ defined as $C \eqdef a.C + \1$;
then, its semantics is an NFA with one state, that is both initial and
final, with an $a$-labeled self-loop; hence, $C$ recognizes the 
language $\{a^n \mid n \geq 0\}$, denotable by the regular expression $a^*$.

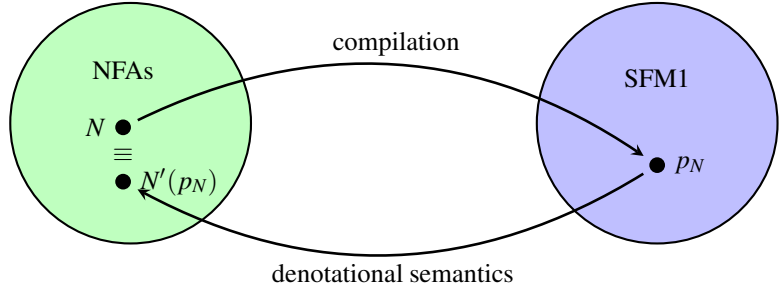
\begin{figure}[h]
\centering
\begin{tikzpicture}
  \tikzset{%
    mythick/.style={%
        line width=.35mm,>=stealth
    }
  }
  \tikzset{%
    mynode/.style={
      circle,
      fill,
      inner sep=2.1pt
    },
    shorten >= 3pt,
    shorten <= 3pt
  }
\def\eodiaglabeldist{0mm}
\def\eolabeldist{0.2mm}
\def\eofigdist{4.5cm}
\def\rrrel{$\cong_r$}
\def\eodist{0.5cm}
\def\eodisty{0.4cm}
\def\eodistw{0.8cm}

\draw [thick,fill=green!25] (-1,-1) circle [radius=1.6cm];
\draw (-1.1,-0.3) node (p0) {NFAs};
\node (p1) [mynode,below =\eodisty of p0, label={[label distance=\eodiaglabeldist]left:$N$}] {};
\node (p2) [mynode,below =\eodist of p1, label={[label distance=\eodiaglabeldist]right:$N'(p_N)$}] {};
 \path (p1) -- node (R3) [inner sep=1pt] {$\equiv$} (p2);

\draw [thick,fill=blue!25] (6,-1) circle [radius=1.6cm];
\draw (6,-0.4) node (q0) {SFM1};
\node (q1) [mynode,below =\eodistw of q0, label={[label distance=\eodiaglabeldist]right:$p_N$}] {};

\draw (p1) edge[mythick,->, bend left] node[above] {compilation} (q1);
\draw (q1) edge[mythick,->, bend left] node[below] {denotational semantics} (p2);

\end{tikzpicture}
\caption{Graphical description of the representability theorem, up to isomorphism}
\label{rep2}
\end{figure}

For SFM1, we get
a {\em Representability Theorem} (see Figure \ref{rep2}) stating that SFM1 truly represents
the class of NFAs, up to isomorphism $\equiv$. In fact,
\begin{itemize}
\item[$(i)$] Given an NFA $N$, we compile it to an SFM1 term representation $p_N$ by an algorithm described in the proof of Theorem \ref{representability-nfa}: so {\em all} NFAs can be represented 
by SFM1; and
\item[$(ii)$] to each SFM1 term $p$ (hence also to $p_N$), we associate, by means of a denotational semantics 
described in Table \ref{den-nfa-sfm1},  an NFA $N'(p)$ (hence also $N'(p_N)$): 
so {\em only} NFAs can be 
represented by SFM1; and, finally,
\item[$(iii)$] the NFAs $N$ and $N'(p_N)$ are isomorphic, denoted by $N \equiv N'(p_N)$.
\end{itemize}

One advantage of this more concrete representability theorem is that the algebra SFM1 can be used as a basis to study
all the possible equivalences that can be defined over NFAs, not only language equivalence $\sim$ (as for regular expressions), but also more concrete equivalences, such as bisimulation equivalence \cite{Park81,Mil89,BBR09,GV15}, or any in the 
linear-time/branching-time spectrum \cite{vG01,GV15}.

The main contribution of this paper is the proof that
language equivalence can be characterized over SFM1 terms by means of a very concise axiomatization, 
comprising only 7 axioms (4 for disjunction and 3 for prefixing) and 3 conditional axioms (for recursion). 
The axioms we present are very similar to known axioms \cite{Kleene,Salomaa,Kozen,Mil84,Mil89a,Gor20ic}.
In fact, for disjunction they are the usual ones: the choice operator is an idempotent, commutative monoid, with $\nil$ as neutral element.
The axioms for prefixing are similar to those for concatenation: $\nil$ is an annihilator, the prefixing operator distributes over a choice term,
and the prefix $\epsilon$ can be absorbed. The axioms for recursion are also similar to axioms developed for SFM \cite{Gor19,Gor20ic}, in turn 
inspired to those for finite-state CCS \cite{Mil89a,Mil89}: the {\em unfolding} axiom (stating that a constant $C$, defined as $C \eqdef p\{C/x\}$, 
can be equated to its body $p\{C/x\}$), the {\em folding} axiom (stating, roughly, that if $C \eqdef p\{C/x\}$ and 
$q = p\{q/x\}$, then $C = q$), and the
{\em excision} axiom (stating that if $C \eqdef (\epsilon.x +q)\{C/x\}$ and $D \eqdef q\{D/x\}$, then $C = D$).

The paper is organized as follows. Section \ref{def-sec} recalls the basic definitions about NFAs, including language equivalence $\sim$, bisimulation equivalence $\simeq$ and isomorphism equivalence $\equiv$. 
Section \ref{sfm1-sec} introduces the algebra SFM1 (i.e., its syntax and its denotational semantics 
in terms of NFAs) and proves the representability theorem sketched in Figure \ref{rep2}.
Section \ref{cong-sec} shows that language equivalence $\sim$ is a congruence for all the operators of SFM1, notably recursion.
Section \ref{alg-prop-sec} studies the algebraic properties of language equivalence over SFM1 terms, that 
are useful to prove the soundness of the axiomatization. A particular care is devoted to prove, by means of 
the theory of fixed points \cite{DP02,San10}, the folding law for recursion, which holds only if 
the recursively defined constant $C$ is {\em observationally guarded}, i.e., cannot perform 
an $\epsilon$-labeled loop: $C \NDeriv{\epsilon} C$.
In Section \ref{eq-ded-sec}, we describe how equational deduction is performed in an algebra using 
process constants for recursion. In Section \ref{axiom-sec}, we first
present the set of axioms and prove that the axiomatization is sound.
Furthermore, we show that the axiomatization is complete; we prove this result first for observationally guarded SFM1 terms 
(by first turning an observationally guarded term into an $\epsilon$-free normal form and then proving the completeness 
for these normal forms using only 9 axioms),
and then we extend the completeness proof also to observationally unguarded SFM1 terms, by showing that,
by means of the additional conditional recursion axiom of excision, each observationally unguarded term can be equated
to an observationally guarded one.
Section \ref{conc-sec} adds concluding remarks and hints for future research.

%
\section{Nondeterministic Finite Automata} \label{def-sec}
%

Let $A$ denote a finite alphabet, ranged over by $a, b, \ldots$. Let $A^*$ denote the set of 
all the words over $A$ (including the empty word
$\epsilon$), ranged over by $\sigma$. 
Let $\alpha$ range over $A \cup \{\epsilon\}$. Let $\mathcal{P}(A^*)$ denote the set of all formal languages
over $A$, ranged over by $L$, possibly indexed.

\begin{definition} A nondeterministic finite automaton (NFA for short) is a tuple $N = (Q, A,$ $T, F, q_0) $ where
	\begin{itemize}
		\item $Q$ is a finite set of states, ranged over by $q$ (possibly indexed);
		\item $A$ is a finite alphabet;
		\item $T \subseteq Q \times (A\cup \{\epsilon\}) \times Q $ is the set of transitions, ranged over by $t$ (possibly indexed);
		\item $F \subseteq Q$ is the set of final states;
		\item $q_0 \in Q$ is the initial state.
	\end{itemize}
Given a transition $ t = (q,\alpha,q')$, $ q $ is called the \textit{source}, $ \alpha $ the \textit{label} of the transition (denoted also by $l(t)$),  and 
$ q' $ the \textit{target}; this is usually represented as $q \deriv{\alpha} q'$. An NFA $N = (Q, A, T, F, q_0) $ is {\em deterministic} if:
\begin{itemize}
\item $\forall t \in T \; l(t) \in A$ (i.e., there are no $\epsilon$ labeled transitions), and
\item $\forall q\in Q \, \forall a \in A$ there exists exactly one state $q' \in Q$ such that $(q, a, q') \in T$, i.e., $T$ can be equivalently represented as a function of type $Q \times A \rightarrow Q$.
\end{itemize}
A deterministic NFA is called a {\em deterministic finite automaton} - DFA, for short.
\fine
\end{definition}

\begin{definition}\label{def-derivstar}	\textbf{(Reachability relation and reduced NFA)}
Given $N = (Q, A, T, F, q_0)$,  the {\em reachability relation} ${\Rightarrow } \subseteq Q \times A^* \times Q$ is the least relation 
	induced by the following axiom and rules:
	
	$\begin{array}{llllllllll}
	\bigfrac{}{ q \Deriv{\epsilon} q} &\quad & \bigfrac{q \deriv{\alpha} q'}{ q \Deriv{\alpha} q'} & \quad & 
	 \bigfrac{q_1 \Deriv{\sigma} q_2 \; \; \; q_2 \deriv{\alpha} q_3 }{q_1 \Deriv{\sigma \alpha} q_3} \\
	\end{array}$\\
	\noindent
	where, of course, we assume that $\epsilon$ is the identity of concatenation: $\sigma \epsilon = \sigma = \epsilon \sigma$.
	We simply write $q_1 \Rightarrow q_2$ to state  that $q_2$ is  {\em reachable} from $q_1$ when
	there exists a word $\sigma$ such that $q_1 \Deriv{\sigma} q_2$.
	We can also define $reach(q)$ as follows: $reach(q) = \{q' \mid q \Rightarrow q'\}$.
	An NFA $N = (Q, A, T, F, q_0)$ is {\em reduced} if $reach(q_0) = Q$.
	\fine
\end{definition}

\begin{figure}[t]
\centering
    \begin{tikzpicture}[shorten >=1pt,node distance=2cm,on grid,auto]
 
    
       \node[state,label={below:$q_0$}]            (q0)               {};
      \node[accepting,state,label={below:$q_1$}] (q1) [right of=q0] {};

      \path[->] (q0) edge               node {$\epsilon$} (q1)
                     edge  [loop above] node {$a$}        (q0)
                (q1) edge  [loop above] node {$b$}        (q1);
       \node[accepting,state,label={below:$q_2$}] (q2)   [right of=q1]            {};
      \node[accepting,state, label={below:$q_3$}] (q3) [right of=q2] {};
      \node[state,label={below:$q_4$}]           (q4) [right of=q3] {};
    
      \path[->] (q2) edge [loop above] node {$a$} (q2)
                     edge              node {$b$} (q3)
                (q3) edge [loop above] node {$b$} (q3)
                     edge              node {$a$} (q4)
                (q4) edge [loop above] node {$a, b$} (q4);

    \end{tikzpicture}    
\caption{An NFA and a DFA both recognizing the language $a^*b^*$}
\label{fig-nfa-rep1}
\end{figure}
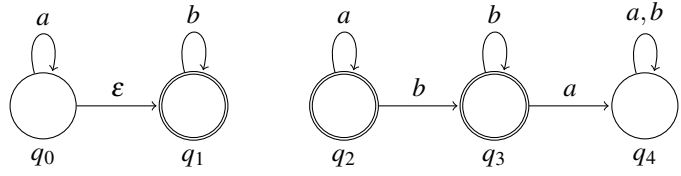

\begin{definition}\label{def-rec-lang}	\textbf{(Recognized language and equivalence)}
Given $N = (Q, A, T, $ $F, q_0)$, the language recognized by $N$ is $L[N] = \{\sigma \in A^* \mid \exists q \in F. q_0 \Deriv{\sigma} q\}$.
The language recognized by state $q \in Q$ is $L[N,q] = \{\sigma \in A^* \mid \exists q' \in F. q \Deriv{\sigma} q'\}$.
Two states $q_1$ and $q_2$ are  equivalent, denoted $q_1 \sim q_2$, if $L[N,q_1] = L[N,q_2]$.
Two NFAs $N_1 = (Q_1, A_1, T_1, F_1, q_{01})$ and $N_2 = (Q_2, A_2, T_2, F_2, q_{02})$ are equivalent if they recognize the same 
language, i.e., $L[N_1] = L[N_2]$.
\fine
\end{definition}

It is well-known that the class of languages recognized by NFAs is the class of {\em regular languages} (see, e.g., \cite{HMU01} for an introductory textbook), i.e., the languages denotable by {\em regular expressions} (originally proposed by Kleene in \cite{Kleene}; we assume the reader is familiar with this algebra).

\begin{example}
Figure \ref{fig-nfa-rep1} outlines two finite automata: the left one is an NFA with initial state $q_0$ and final state $q_1$, and the right one is a DFA 
with initial state $q_2$ and final states $q_2$ and $q_3$. Both recognize 
the same regular language $\{a^nb^m \mid n,m \geq 0\}$, expressible by the regular expression $a^*b^*$.
As a matter of fact, $q_0 \sim q_2$ because $L[N,q_0] = L[N,q_2]$, where $N$ is the union of the two automata (union that can be done 
because the sets of states are disjoint).
\fine
\end{example}

More concrete equivalences can be defined on NFAs, e.g., bisimulation equivalence \cite{Park81,Mil89,BBR09,GV15} and isomorphism equivalence.

\begin{definition}\label{def-bisf}{\bf (Bisimulation)}
Let $N = (Q, A, T, F, q_0)$ be an NFA. A {\em bisimulation} is a relation
$R\subseteq Q\times Q$ such that if $(q_1, q_2) \in R$, then for all $\alpha \in A\cup\{\epsilon\} $
	\begin{itemize}
		\item $ \forall q_1' $ such that $ q_1 \deriv{\alpha} q_1' $, $ \exists q_2' $ such 
		that $ q_2 \deriv{\alpha} q_2' $ and $ (q_1', q_2') \in R $,
		\item $ \forall q_2' $ such that $ q_2 \deriv{\alpha} q_2' $, $ \exists q_1' $ such that 
		$ q_1 \deriv{\alpha} q_1' $ and $ (q_1', q_2') \in R $, and
                 \item $q_1 \in F$ iff $q_2 \in F$.
       \end{itemize}
Two states $q$ and $q'$ are bisimilar, denoted $q \simeq q'$, if there exists a bisimulation $R$ such that $(q, q') \in R$.
\fine
\end{definition}

\begin{example}
Given an NFA $N = (Q, A, T, F, q_0)$, it is easy to see that if $q_1 \simeq q_2$ then $q_1 \sim q_2$. 
The converse implication does not hold. For instance, consider Figure \ref{fig-nfa-rep1}: of 
course, $q_0 \sim q_2$, but $q_0 \not\simeq q_2$.
\fine
\end{example}

\begin{remark}\label{lang=bis-rem}
It is interesting to observe that if $N = (Q, A, T, F, q_0)$ is a DFA, then language equivalence coincides with bisimulation equivalence.
As a matter of fact, it is easy to see that for DFAs
the relation $R = \{(q_1, q_2) \mid q_1 \sim q_2\}$ is a bisimulation.
\fine
\end{remark}

\begin{definition}\label{def-iso-lts}\index{Isomorphism}\textbf{(Isomorphism)}
Two NFAs $N_1 = (Q_1, A_1, T_1, F_1, q_{01})$ and $N_2 = (Q_2, A_2, T_2,$ $ F_2, q_{02})$ are {\em isomorphic}, denoted by
$N_1 \equiv N_2$,
if there exists a bijection $f: Q_1\rightarrow Q_2$ such that:
\begin{itemize}
\item it is type-preserving: $q_1 \in F_1$ iff $f(q_1) \in F_2$;
\item it preserves transitions: 
$
q\derivu{\alpha} q'  \mbox{  iff  }   f(q)\derivd{\alpha} f(q')
$ 
for all $q, q' \in Q_1$ and for all $\alpha \in A_1 \cup A_2 \cup \{\epsilon\}$; and
\item it preserves the initial states: $f(q_{01}) = q_{02}$.\\[-.9cm]
\end{itemize}
\fine
\end{definition}

%
\section{SFM1: Syntax and Semantics}\label{sfm1-sec}
%

The process algebra SFM1 we are going to define is a slight extension to a dialect of finite-state CCS \cite{Mil89,GV15}, 
called SFM in \cite{Gor17}, where an additional constant $\1$ is introduced in order to distinguish final states from non-final ones, as introduced in
BPA and related languages (see, e.g., \cite{BBR09} for an extensive investigation of this issue). 

%
\subsection{Syntax}\label{sfm1-syn-sec}
%

Let $\cons$ be a finite set of constants, disjoint from the alphabet
$A$, ranged over by $C, D, \ldots$ (possibly indexed). The size of the finite sets $A$ and $\cons$ is not important: 
we assume that they can be chosen as large as needed.  
The SFM1 {\em terms} 
are generated from alphabet symbols and constants by the 
following abstract syntax:\\

$\begin{array}{llllllllllllll}
s &  ::= &   \1 & | & \nil & | & \alpha.p & | &  s+s &  \hspace{1 cm} \mbox{{\em guarded processes}}\\
p & ::= & s & | & C & &&&&\hspace{1 cm} \mbox{{\em  processes}}\\
\end{array}$\\

\noindent
where the special constants $\1$ and $\nil$ denote, respectively, a {\em final} state and a {\em non-final} state,
$\alpha.p$ is a process where
the symbol $\alpha \in A \cup \{\epsilon\}$ prefixes the residual $p$ ($\alpha.-$ is a family of {\em prefixing} operators, one for each $\alpha$),
$s_1 + s_2$ denotes the disjunction of $s_1$ and $s_2$ ($- + -$ is
the {\em choice} operator), and $C$ is a constant. 
A constant $C$ may be equipped with a definition, but this must be a guarded process, i.e., 
$C \eqdef s$. A term $p$ is {\em final} if $p \downarrow$ holds, where 
the predicate $\downarrow$ is the least one generated by the axiom and the rules outlined in Table \ref{downarrow-rules}.

By \const{p} we denote the set of process constants $\delta(p, \emptyset)$ used by $p$, where the
auxiliary function $\delta$, which has, as an additional parameter, a set $I$ of already known constants, is 
defined as follows:\\

 $\begin{array}{rcllrclrcl}
\delta(\1, I) & = & \emptyset & \quad &
\delta(\nil, I) & = & \emptyset \\
\delta(\alpha.p, I) & = & \delta(p, I) & \quad &
\delta(p_1 + p_2, I) & = & \delta(p_1, I) \cup \delta(p_2, I)\\ 
\end{array}$\\
 $\begin{array}{rclrclrcl}
\quad \delta(C, I) & = & \begin{cases}
  \emptyset & \! \! \mbox{$C \in I $,}\\  
   \{C\} & \! \! \mbox{$C \not\in I  \wedge C$ undefined,} \\
  \{C\} \cup  \delta(p, I \cup \{C\})  & \! \! \mbox{$C \not\in I  \wedge C \eqdef p$.} \\
   \end{cases}
\end{array}$\\

A term $p$ is a SFM1 {\em process} if  \const{p} is finite and each constant in \const{p} is equipped 
with a defining equation (in category $s$).
The set of SFM1 processes is denoted by $\mathcal{P}_{SFM1}$, while the set of its guarded processes, i.e.,
those in syntactic category $s$, by $\mathcal{P}_{SFM1}^{grd}$.

\begin{table}[t]
{\renewcommand{\arraystretch}{3}
\hrulefill\\[-1.3cm]
\normalsize{

\begin{center}

$\begin{array}{llllllllllll}
\bigfrac{}{\1 \downarrow} & \qquad \quad & \bigfrac{p_1 \downarrow}{(p_1 + p_2)\downarrow} &  \qquad \quad & \bigfrac{p_2 \downarrow}{(p_1 + p_2)\downarrow} & \qquad \quad &   \bigfrac{p \downarrow \; \; C \eqdef p}{C \downarrow}
\end{array}$\\

\hrulefill\\
\end{center}}}
\caption{Final states: predicate $\downarrow$}\label{downarrow-rules}
\end{table}

%
\subsection{Denotational Semantics}\label{sfm1-den-sec}
%

Now we provide a construction of the NFA $\encodings{\emptyset}{p}$
associated with process $p$, which is compositional and denotational in style. The details of
the construction are outlined in Table \ref{den-nfa-sfm1}. The encoding is parametrized by a set of constants
that has  already been found while scanning $p$; such a set is initially empty and it is used to avoid 
looping on recursive constants. The definition is syntax driven
and also the states of the constructed NFA are syntactic objects, i.e., SFM1 process terms. 
A bit of care is needed in the rule for choice: in order to include only strictly necessary 
states and transitions, the initial state $p_1$ (or $p_2$) of the NFA $\encodings{I}{p_1}$ (or $\encodings{I}{p_2}$) 
is to be kept in the NFA for $p_1 + p_2$ only if there exists a transition reaching the state $p_1$ (or $p_2$) in $\encodings{I}{p_1}$
(or $\encodings{I}{p_2}$), 
otherwise $p_1$ (or $p_2$) can be safely removed in the new NFA.
Similarly, for the rule for constants. In Table \ref{den-nfa-sfm1} we denote 
by $T(q) = \{t \in T \mid \exists \alpha \in A \cup \{\epsilon\}, \exists q' \in Q. t = (q, \alpha, q')\}$  
the set of transitions in $T$ with $q$ as source state.

\begin{table}[t]
	{\renewcommand{\arraystretch}{1}
		\hrulefill\\[-1.1cm]
		
		\begin{center}\fontsize{8pt}{-0.1pt}
			$\begin{array}{rcllcllcl}
			\encodings{I}{\1}  & =  & (\{\1\}, \emptyset, \emptyset, \{\1\}, \1) &\\
			\encodings{I}{\nil}  & =  & (\{\nil\}, \emptyset, \emptyset, \emptyset, \nil) &\\
			\encodings{I}{\alpha.p}  & =  & (Q, A, T, F, \alpha.p) & \mbox{given } 
			\encodings{I}{p}   =   (Q', A', T', F', p) \; \mbox{ and where } Q = \{\alpha.p\} \cup Q' \\ 
			& & & F = F', \,T = \{(\alpha.p, \alpha, p)\} \cup T', \,
			A = \begin{cases}
			\{\alpha\} \cup A' \hspace {1em} \mbox{ if $\alpha \neq \epsilon$}\\  
			A' \hspace{4em} \mbox{o.w.} \\
			\end{cases}\\	
			\encodings{I}{p_1 + p_2}  & =  & (Q, A, T, F, p_1 + p_2)& \mbox{given }  \encodings{I}{p_i}   =   
			(Q_i, A_i, T_i, F_i, p_i) \; \mbox{ for $i = 1, 2$, and where} \\  
			& & &  
			Q = \{p_1 + p_2\} \cup Q_1' \cup Q_2', \mbox{ with, for $i = 1, 2$, } \\
			& & & Q'_i = \begin{cases}
			Q_i \hspace {1em} \mbox{ 
				$\exists t \in T_i$ such that $t = (q, \alpha, p_i)$}\\  
			Q_i \setminus \{p_i\} \hspace{4em} \mbox{o.w.} \\
			\end{cases}\\
			& & &  A = A_1 \cup A_2, \;  T = T' \cup T'_1 \cup T'_2,  \mbox{ with, for $i = 1, 2$, }\\
			& & & T'_i = \begin{cases}
			T_i \hspace {4em} \mbox{
				$\exists t \in T_i. \, t = (q, \alpha, p_i)$}\\  
			T_i \setminus T_i(p_i)  \hspace{2em} \mbox{o.w.} \\
			\end{cases}\\ & & & 
			T' = \{(p_1 + p_2, \alpha, q) \mid (p_i, \alpha, q) \in T_i, i = 1, 2\}\\
			& & & F'_i = F_i \cap Q'_i \mbox{ for $i = 1, 2$, and  } \\
			&&& F = \begin{cases}
			F'_1 \cup F'_2 \cup \{p_1+p_2\} \hspace {1em} \mbox{
				if $p_1 \in F_1$ or $p_2 \in F_2$}\\  
			F'_1 \cup F'_2 \hspace{6em} \mbox{o.w.} \\
			\end{cases}\\ 
			\encodings{I}{C}  & =  & (\{C\}, \emptyset, \emptyset, \emptyset, C) & \mbox{if $C \in I$ } \\
			\encodings{I}{C}  & =  & (Q, A, T, F, C) & \mbox{if $C \not \in I$, given $C \eqdef p$ and } 
			\encodings{I\cup \{C\}}{p}   =   (Q', A', T', F', p) \\\ 
			&&&  A = A', Q =  \{C\} \cup Q'', \mbox{ where }\\
			& & & Q'' =  \begin{cases}
			Q'  \hspace {1em} \mbox{
				$\exists t \in T'$ such that $t = (q, \alpha, p)$}\\  
			Q' \setminus \{p\} \hspace{4em} \mbox{o.w.} \\ 
			\end{cases}\\
			& & & T = \{(C, \alpha, q) \mid (p, \alpha, q) \in T'\} \cup T'' \mbox{ where }\\ & & &
			T'' =   \begin{cases}
			T'  \hspace {4em} \mbox{
				$\exists t \in T'$. $t = (q, \alpha, p)$}\\  
			T' \setminus T'(p) \hspace{4em} \mbox{o.w.} \\ 
			\end{cases}\\ 
			& & & F'' = F' \cap Q'' \mbox{ and  } 
			 F = \begin{cases}
			F'' \cup \{C\} \hspace {1em} \mbox{
				if $p \in F'$ }\\  
			F'' \hspace{4em} \mbox{o.w.} \\
			\end{cases}\\ 		
			\end{array}$
				
			\hrulefill
	\end{center}}
	\caption{Denotational semantics}\label{den-nfa-sfm1}
\end{table}

\begin{figure}[t]
\centering
    \begin{tikzpicture}[shorten >=1pt,node distance=2.8cm,on grid,auto]
    
      \node[state,label={below:$C$}]           (q10)                {};
      \node[state,label={below:$D$}]           (q11) [right of=q10] {};
      \node[accepting,state,label={below:$b.D+\1$}] (q12) [right of=q11] {};
    
      \path[->] (q10) edge              node {$b$} (q11)
                (q11) edge [bend right] node {$a$} (q12)
                (q12) edge [bend right] node {$b$} (q11);  
            
    \end{tikzpicture}    
\caption{The NFA for $C \eqdef b.D$, where $D \eqdef a.(b.D + \1)$, of Example \ref{ex-den-sfm1}}
\label{fig-nfa-den}
\end{figure}
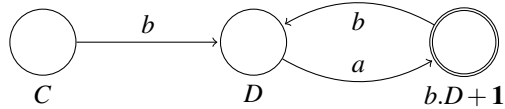

\begin{example}\label{ex-den-sfm1}
Consider constant $C \eqdef b.D$, where $D \eqdef a.(b.D + \1)$. By using the definitions
in Table \ref{den-nfa-sfm1}, $\encodings{\{C,D\}}{D} = (\{D\}, \emptyset, \emptyset, \emptyset, D)$. Then, by prefixing,

$\encodings{\{C,D\}}{b.D}$ $ = (\{b.D, D\}, \{b\}, \{(b.D, b, D)\}, \emptyset, b.D)$. 

Now, $\encodings{\{C,D\}}{\1}$ $ = (\{\1\}, \emptyset, \emptyset, \{\1\}, \1)$. Then, by summation

$\encodings{\{C,D\}}{b.D + \1}$ $ = (\{b.D + \1, D\}, \{b\}, \{(b.D +1, b, D)\}, \{b.D + \1\}, b.D +\1)$.

Note that the states $b.D$ and $\1$ have been removed, as no transition in $\encodings{\{C,D\}}{b.D}$ reaches $b.D$ and
no transition in $\encodings{\{C,D\}}{\1}$ 
reaches $\1$.

Again, by prefixing,
$\encodings{\{C,D\}}{a.(b.D+\1)} = (\{a.(b.D+\1), b.D+\1, D\}, \{a, b\},$ 

$ \{(a.(b.D+\1), a, b.D+\1), (b.D+\1, b, D)\}, \{b.D + \1\}, a.(b.D+\1))$.

\noindent
Now, the rule for constants ensures that 

$\encodings{\{C\}}{D} = (\{D, b.D+\1\}, \{a, b\}, \{(D, a, b.D+\1), (b.D+\1, b, D)\}, \{b.D+\1\}, D)$.

\noindent
Note that state $a.(b.D+\1)$ has been removed, as no transition in $\encodings{\{C,D\}}{a.(b.D+\1)}$ reaches that state.
By prefixing, 

$\encodings{\{C\}}{b.D} = (\{b.D, D, b.D+\1\}, \{a, b\}, \{(b.D, b, D), (D, a, b.D+\1),  (b.D+\1, b, D)\},$ $ \{b.D+\1\}, b.D)$, 

\noindent 
Finally, 
$\encodings{\emptyset}{C} = (\{C, D, b.D+\1\}, \{a, b\}, \{(C, b, D), (D, a, b.D+\1), (b.D+\1, b, D)\},$ 

$  \{b.D+\1\}, C)$. The resulting NFA is outlined in Figure \ref{fig-nfa-den}.
\fine
\end{example}

\begin{theorem}\label{finite-reduced-NFA}
For each SFM1 process $p$, $\encodings{\emptyset}{p}$ is a reduced NFA.	
	\proof
	By induction on the definition of $\encodings{I}{p}$. Then, the thesis follows for $I = \emptyset$.
	
	The first base case is for $\nil$ and the thesis is obvious, as,
	for any $I \subseteq \cons$, $ \encodings{I}{\nil} = (\{\nil\}, \emptyset, \emptyset, \emptyset, \nil)$, which is clearly a reduced NFA.
	Similarly for $\1$.
	The third base case is for $C$ when $C \in I$, which corresponds to the case when $C$ is not defined.
	In such a case, for any $I \subseteq \cons$ with $C \in I$, $\encodings{I}{C} =  (\{C\}, \emptyset, \emptyset, \emptyset, C)$, which
	is a reduced NFA.	
	
	The inductive cases are as follows.

	{\em Prefixing}: By induction, we assume that $\encodings{I}{p} = (Q', A', T', F', p)$ is a reduced NFA.
		Hence, $\encodings{I}{\alpha.p} = (Q' \cup \{\alpha.p\}, A, T' \cup \{(\alpha.p, \alpha, p)\}, F', \alpha.p)$ (where $A = A' \cup \{\alpha\}$
		if $\alpha \neq \epsilon$, otherwise $A = A'$),
		is a reduced NFA as well.
	
	{\em Choice}: By induction, we can assume that $\encodings{I}{p_i} =  (Q_i, A_i, T_i, F_i, p_i)$ for $i = 1, 2$, are reduced NFAs.
	Hence, $\encodings{I}{p_1+p_2} = (Q, A, T, F, p_1 + p_2)$ is an NFA as well, because, according to the definition 
	in Table \ref{den-nfa-sfm1}, 
	 $|Q| \leq 1 + |Q_1| + |Q_2|$, $|A| \leq  |A_1| +  |A_2|$ and $|T| \leq |T'| + |T_1| + |T_2|$, 
	 where $|T'| \leq |T_1| + |T_2|$; and also reduced by construction.
	
	{\em Constant}: In this case, we assume that $C \eqdef p$ and that $C \not \in I$.
	By induction, we can assume that $\encodings{I\cup \{C\}}{p} =  (Q', A', T', F', p)$ is a reduced NFA.
	Hence, $\encodings{I}{C} = (Q, A, T, F, C)$ is an NFA as well, because, according to the definition 
	in Table \ref{den-nfa-sfm1}, $|Q| \leq 1 + |Q'|$, $A = A'$ and $|T| \leq 2 \cdot |T'|$; and also reduced by construction.
	\fine
\end{theorem}

\begin{theorem}\label{representability-nfa}
{\bf (Representability)} For any reduced NFA $N$, there exists an SFM1 process $p$ such that $\encodings{\emptyset}{p}$
is isomorphic to $N$. 
\proof
Let $N = (Q, A, T, F, q_0) $ be a reduced NFA, with $Q = \{q_0, q_1, \ldots, q_n\}$. 
For $i = 1, \ldots, n$, let $T(q_i) = \{t \in T \mid \exists \alpha \in A \cup \{\epsilon\}, \exists q_k \in Q. t = (q_i, \alpha, q_k)\}$ be 
the set of transitions in $T$ with $q_i$ as source state. We define a process constant $C_i$ in correspondence
with state $q_i$, for $i = 0, 1, \ldots, n$, defined as follows:
\begin{itemize}
\item if $q_i \not\in F$ and $T(q_i) = \emptyset$ (i.e., $q_i$ is a deadlock), then $C_i \eqdef \nil$; 
\item if $q_i\in F$ and $q_i$ is a deadlock, then $C_i \eqdef \1$;
\item if $q_i \not\in F$ and $T(q_i) \neq \emptyset$, then $C_i \eqdef\Sigma_{(q_i, \alpha, q_k) \in T(q_i)} \alpha.C_k$;
\item if $q_i \in F$ and $T(q_i) \neq \emptyset$, then $C_i \eqdef\Sigma_{(q_i, \alpha, q_k) \in T(q_i)} \alpha.C_k +\1$.
\end{itemize}
Let us consider $\encodings{\emptyset}{C_0}$. It is not difficult to see
that $reach(C_0) = \{C_0, C_1 \ldots, C_n \}$ because $N$ is reduced. Hence, the bijection we are looking for is
$f: Q \rightarrow \{C_0, C_1 \ldots, C_n \}$, defined as $f(q_i) = C_i$.
It is also easy to observe that the three conditions of isomorphism are satisfied, namely:
\begin{itemize}
\item $q \in F\;$ iff $\; f(q)\downarrow$, and 
\item $q\derivu{\alpha}q' \in N \;$ iff $ \; f(q)  \derivd{\alpha}  f(q') \in \encodings{\emptyset}{C_0}$, and
\item $C_{0} = f(q_{0})$. 
\end{itemize}
Hence, $f$ is indeed an NFA isomorphism.
\fine
\end{theorem}

\begin{corollary}\label{cor-rep}
SFM1 represents, up to isomorphism, the class of reduced NFAs.
\proof
By Theorem \ref{finite-reduced-NFA} the semantics of an SFM1 process is a reduced NFA; hence, only reduced NFAs can be represented by SFM1 processes. By Theorem \ref{representability-nfa} each reduced NFA
 can be represented, up to isomorphism, by a suitable SFM1 process; hence, all reduced NFAs can be represented by SFM1 processes.
\fine
\end{corollary}

\begin{figure}[t]
\centering
    \begin{tikzpicture}[shorten >=1pt,node distance=2cm,on grid,auto]

 
      \node[state,label={below:$q_5$}]           (q5)             {};
      \node[accepting,state,label={below:$q_6$}] (q6) [right of=q5] {};
    
      \path[->] (q5) edge              node {$a$} (q6)
                     edge [loop above] node {$b$} (q5);
                 

     \node[state,label={below:$q_7$}]           (q7)     [right of=q6]      {};
      \node[accepting,state,label={above:$q_9$}] (q9) [below right of=q7] {};
      \node[state,label={below:$q_8$}]           (q8) [above right of=q9] {};
    
      \path[->] (q7) edge              node {$a$} (q8)
                     edge              node {$b$} (q9)
                (q9) edge              node {$a$} (q8)
                     edge [loop below] node {$b$} (q9);

    \end{tikzpicture}    

\caption{Two NFAs for Example \ref{rep-ex}}
\label{fig-nfa-rep2}
\end{figure}
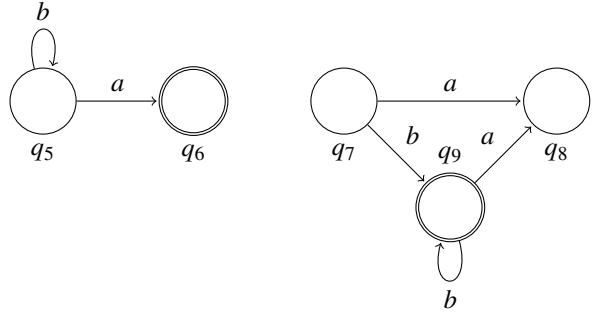

\begin{example}\label{rep-ex}
Let us consider the NFA in Figure \ref{fig-nfa-rep1}. According to the construction in the proof of Theorem \ref{representability-nfa}, the SFM1 representation is as follows:

$\begin{array}{rcl}
C_0 & \eqdef & a.C_0 + \epsilon.C_1\\
C_1 & \eqdef & b.C_1 + \1\\
\end{array}$

\noindent
If we consider the DFA in Figure \ref{fig-nfa-rep1}, the SFM1 representation is as follows:

$\begin{array}{rcl}
C_2 & \eqdef & a.C_2 + b.C_3 + \1\\
C_3 & \eqdef & b.C_3 + a.C_4 + \1\\
C_4 & \eqdef & a.C_4 + b.C_4\\
\end{array}$

Note that if the NFA is not reduced, then we cannot find an SFM1 process that represents it. For instance, if we take 
the union of the two automata in Figure \ref{fig-nfa-rep1} with initial state $q_0$, then $C_0$ represents only 
the NFA on the left.

\noindent
If we consider the first NFA in Figure \ref{fig-nfa-rep2}, we get:

$\begin{array}{rcl}
C_5 & \eqdef & a.C_6 + b.C_5\\
C_6 & \eqdef & \1\\
\end{array}$

\noindent
Finally, if we consider the second NFA in Figure \ref{fig-nfa-rep2}, we get:

$\begin{array}{rcl}
C_7 & \eqdef & a.C_8 + b.C_9\\
C_8 & \eqdef & \nil\\
C_9 & \eqdef & a.C_8 + b.C_9 + \1\\
\end{array}$\\[-.4cm]

\fine
\end{example}

%
\section{Congruence}\label{cong-sec}
%

In the following, for each SFM1 process $p$, its associated language $L(p)$ is the language recognized by
the reduced NFA $\encodings{\emptyset}{p}$, computed according to the denotational semantics 
in Table \ref{den-nfa-sfm1}.

\begin{definition}
Two SFM1 processes $p$ and $q$ are {\em language equivalent}, denoted $p \sim q$, if
$L(p) = L(q)$, i.e., they recognize the same language.
\fine
\end{definition}

\begin{proposition}\label{prop-obvious} The following hold:

\noindent
1) $L(\nil) = \emptyset$.

\noindent
2) $L(\1) = \{\epsilon\}$.

\noindent
3) For each $p \in  \mathcal{P}_{SFM1}$, for all $a \in A$, we have that 
$L(a.p) = \{a w \mid w \in L(p)\}$.

\noindent
4) For each $p \in  \mathcal{P}_{SFM1}$, we have that 
$L(\epsilon.p) =  \{\epsilon w \mid w \in L(p)\} = L(p)$.

\noindent
5) For each $p, q \in  \mathcal{P}_{SFM1}^{grd}$, $L(p+q) = L(p) \cup L(q)$.

\proof Obvious by construction in Table \ref{den-nfa-sfm1}.
\fine
\end{proposition}

Now we show that language equivalence is a congruence for all the SFM1 operators.

\begin{proposition}\label{prop-cong1}
1) For each $p, q \in  \mathcal{P}_{SFM1}$, if $p \sim q$, then $\alpha.p  \sim  \alpha.q$
for all $\alpha \in A \cup \{\epsilon\}$.  

\noindent
2) For each $p, q \in  \mathcal{P}_{SFM1}^{grd}$, if $p \sim q$, 
then $p + r  \sim  q + r \;$
for all $r \in  \mathcal{P}_{SFM1}^{grd}$.

\proof By hypothesis, we know that $L(p) = L(q)$. Then the thesis follows easily by application of Proposition \ref{prop-obvious}.

Now, for case 1, $L(\alpha.p) = \{\alpha w \mid w \in L(p)\} = \{\alpha w \mid w \in L(q)\} = L(\alpha.q)$ and so
$\alpha.p  \sim  \alpha.q$.
While, for case 2, $L(p+r) = L(p)\cup L(r) = L(q) \cup L(r) = L(q+r)$ and so  
$p + r  \sim  q + r$.
\fine
\end{proposition}

Note that the symmetric case $r + p \sim r + q$ is implied by the fact that 
the operator of choice is commutative w.r.t. $\sim$ (see Proposition \ref{aci-+}).

Still there is one construct missing: recursion, defined over guarded terms only.
Consider an extension of SFM1 where terms can be constructed using variables, such as $x, y, \ldots$
(which are in syntactic category $p$): 
this defines
an ``open'' SFM1 that has the following syntax:\\

$\begin{array}{lccccccccccl}
s &  ::= & \1 & | &  \nil & | & a.p & | &  s+s & \hspace{1 cm}\\
p & ::= & s & | & C & | & x &&\hspace{1 cm}\\
\end{array}$\\

\noindent
where the semantic of $x$ is the NFA $(\{x\}, \emptyset, \emptyset, \emptyset, x)$.

Sometimes we use the notation $p(x_1, \ldots, x_n)$ to state explicitly that term $p$ is open on the tuple of variables $(x_1, \ldots, x_n)$.
For instance, $p_1(x) = a.(b.\1 + c.x)$ and $p_2(x) = a.c.x + a.b.\1$ are open guarded SFM1 terms.
Similarly, $C_1(x) \eqdef a.x$ denotes explicitly that the constant $C_1$ is defined by a term open on $x$.

An open term $p(x_1, \ldots, x_n)$ can be {\em closed} by means of a substitution as follows (see also Section \ref{eq-ded-sec} for the definition of substitution application, also in the case of process constants):
\[
p(x_1, \ldots, x_n)\{r_1 / x_1, \ldots, r_n / x_n\}
\]
\noindent
with the effect that each occurrence of the free variable $x_i$ is replaced by the {\em closed} SFM1
process $r_i$, for $i = 1, \ldots, n$. For instance,
$p_1(x)\{d.\nil / x\} = a.(b.\1 + c.d.\nil)$, as well as $C_1(x)\{d.\nil / x\}$ is $C_1 \eqdef a.d.\nil$ (i.e., we close the open
term $a.x$ defining the body of $C_1$).
A natural extension of language equivalence $\sim$ over open {\em guarded} terms is as follows:
$p(x_1, \ldots, x_n) \sim q(x_1, \ldots, x_n)$
if for all tuples of closed SFM1 terms
$(r_1, \ldots, r_n)$, \\

$p(x_1, \ldots, x_n)\{r_1 / x_1, \ldots, r_n / x_n\}$ $\sim$ $q(x_1, \ldots, x_n)\{r_1 / x_1, \ldots, r_n / x_n\}$.\\

\noindent
E.g., it is easy to see that $p_1(x) \sim p_2(x)$. As a matter of fact, for all $r$, 
$p_1(x)\{r / x\} = a.(b.\1 + c.r)$ $\sim$ $a.c.r + a.b.\1$ = $p_2(x)\{r / x\}$, which can be easily 
proved by means of the algebraic properties (see the next section) and the congruence ones of $\sim$.

For simplicity's sake, let us now restrict our attention to open guarded terms using a single undefined variable.
We can {\em recursively close} an open term $p(x)$ by means of a recursively defined constant. For instance, 
$C \eqdef p(x)\{C / x\}$. The resulting process constant $C$ is a closed SFM1 process. 
By saying that language equivalence is a congruence for
recursion we mean the following: If $p(x) \sim q(x)$ and $C \eqdef p(x)\{C / x\}$ and $D \eqdef q(x)\{D / x\}$, 
then $C \sim D$.
The following theorem states this fact. 
For simplicity's sake, in the following a term $p$, open on a variable $x$, is annotated 
as $p(x)$ only when it is unclear from the context, because we work with real terms and not with annotated ones.

\begin{remark}
Throughout the paper, it is assumed that whenever a constant $C$ is defined by $C \eqdef p\{C/x\}$,
then the open guarded term $p$ does not 
contain occurrences of $C$, i.e., $C \not \in \const{p}$,
so that all the instances of $C$ in $p\{C/x\}$ are due to substitution of $C$ for $x$. Moreover, it is assumed that for each constant
definition we use a different variable, for instance, $D \eqdef q\{D/y\}$.
\fine
\end{remark}

\begin{theorem}\label{recurs-cong-th}\index{Congruence}
	Let $p$ and $q$ be two open guarded SFM1 terms, with one variable $x$ at most. 
	Let $C \eqdef p\{C / x\}$, $D \eqdef q\{D / x\}$
	and $p \sim q$. Then $C \sim D$.
	
	\proof For the term $p$, open on $x$, we define the following languages:
	\begin{itemize}
	\item $L_p^{\downarrow}= \{\sigma \in A^* \mid \exists p' \in F. p \Deriv{\sigma} p'\}$ is the set of the words recognized by $p$;
         \item $L_p^x= \{\sigma \in A^* \mid p \Deriv{\sigma} x\}$ is the set of its maximal words reaching the variable $x$.
  	\end{itemize}
It is rather easy to see that for $C \eqdef p\{C / x\}$, we have that  $L(C) = (L_p^x)^* \cdot L_p^{\downarrow}$, 
where $\_ \cdot \_$ is the concatenation operator and $\_^*$ is the Kleene star operator over languages \cite{HMU01}.

	It is not difficult to see that $p \sim q$ if and only if $L_p^{\downarrow} = L_q^{\downarrow}$ and $L_p^x = L_q^x$, from which the thesis $C \sim D$ follows immediately.
		\fine
\end{theorem}

\begin{example}
Consider $p_1(x) = a.(b.\1 + c.x)$ and $p_2(x) = a.c.x + a.b.\1$. Of course, they are language equivalent and indeed
$L_{p_1}^{\downarrow} = \{ab\} = L_{p_2}^{\downarrow}$ and $L_{p_1}^x = \{ac\} = L_{p_2}^x$. 
If we consider $C \eqdef p_1\{C / x\}$ and
$D \eqdef p_2\{D / x\}$, then we get that $C \sim D$, as both recognize the regular language $(ac)^*ab$.
\fine
\end{example}

The generalization to open terms over a set $\{x_1, \ldots, x_n\}$ of variables is rather obvious. 
Consider a set of pairs of equivalent open terms of this form:\\

$\begin{array}{lrcllllll}
\mbox{$p_1(x_1, \ldots, x_n) \sim q_1(x_1, \ldots, x_n) $}\\
\mbox{$p_2(x_1, \ldots, x_n) \sim q_2(x_1, \ldots, x_n) $}\\
\mbox{$\ldots$}\\
\mbox{$p_n(x_1, \ldots, x_n) \sim q_n(x_1, \ldots, x_n) $}\\
\end{array}$\\

\noindent
where, for $i = 1, \ldots, n$, $p_i(x_1, \ldots, x_n) \sim q_i(x_1, \ldots, x_n)$ if and only if $L_{p_i}^{\downarrow} = L_{q_i}^{\downarrow}$
and $L_{p_i}^{x_j} = L_{q_i}^{x_j}$ for $j = 1, \ldots, n$.
Then, we can recursively close these terms as follows:\\

$\begin{array}{lrcllllll}
\mbox{$C_1 \eqdef p_1\{C_1/x_1, \ldots, C_n/x_n\} \hspace {3em} D_1 \eqdef q_1\{D_1/x_1, \ldots, D_n/x_n\}$}\\
\mbox{$C_2 \eqdef p_2\{C_1/x_1, \ldots, C_n/x_n\} \hspace {3em} D_2 \eqdef q_2\{D_1/x_1, \ldots, D_n/x_n\}$}\\
\mbox{$\ldots$}\\
\mbox{$C_n \eqdef p_n\{C_1/x_1, \ldots, C_n/x_n\} \hspace {3em} D_n \eqdef q_n\{D_1/x_1, \ldots, D_n/x_n\}$}\\
\end{array}$\\

\noindent
Then, the thesis is that $C_i \sim D_i$ for all $i = 1, 2, \ldots, n$.

Finally, a comment on the more concrete equivalences we have defined in Section \ref{def-sec}. Isomorphism equivalence is the most concrete
and it is not a congruence for the choice operator. For instance, $a.\1$ generates an NFA isomorphic to the one for $a.(\1+\nil)$; however,
if we consider the context $\_ + a.\1$ we get two terms generating not isomorphic NFA: $a.\1 + a.\1$ generates a two-state NFA, while
$a.(\1 + \nil) + a.\1$ generates a three-state NFA.
On the contrary, it is possible to prove that bisimulation equivalence $\simeq$, which is strictly finer than language equivalence 
$\sim$, is a congruence for all the operators of SFM1, by means of standard proof techniques (see, e.g., \cite{Mil89,GV15,BBR09,San10,Gor20ic}).

%
\section{Algebraic Properties}\label{alg-prop-sec}
%

Now we list some algebraic properties of language equivalence and discuss whether they hold for bisimulation equivalence or isomorphism equivalence.

\begin{proposition}\label{aci-+}{\bf (Laws of the choice op.)}
For each $p, q, r \in  \mathcal{P}_{SFM1}^{grd}$ , the following hold:
 
 $\begin{array}{lrcllllll}
\qquad & p + (q + r)  &\sim & (p + q) + r & \quad &\; \mbox{(associativity)}\\
\qquad & p + q  &\sim & q + p & \quad &\;  \mbox{(commutativity)}\\
\qquad & p + \nil  &\sim & p &  &\;  \mbox{(identity)}\\
\qquad & p + p  &\sim & p &  & \; \mbox{(idempotency)}
\end{array}$
\proof The proof follows directly from Proposition \ref{prop-obvious}(5). For instance,
$p + q  \sim  q + p$ holds because $L(p+q) = L(p) \cup L(q) = L(q) \cup L(p) = L(q+p)$.
\fine
 \end{proposition}
 
 The laws of the choice operator hold also for bisimilarity \cite{Mil89,GV15,BBR09,San10},
 but they do not hold for isomorphism equivalence. For instance, consider $p = a.A$, with $A \eqdef a.(b.B + a.A)$,
 and $q = b.B$, with $B \eqdef b.B + \1$; then, it is easy to see that the NFA for $p + q$ (four states) 
 is not isomorphic to the NFA for $q + p$ (three states). 
 
 \begin{proposition}\label{pref-alg-prop}{\bf (Laws of the prefixing operator)}
 For each $p, q \in  \mathcal{P}_{SFM1}^{grd}$, for each $r \in \mathcal{P}_{SFM1}$,
 for each $\alpha \in A \cup \{\epsilon\}$, the following hold:
 
 $\begin{array}{lrcllllll}
\qquad & \alpha.\nil  &\sim & \nil & \quad &\; \mbox{(annihilation)}\\
\qquad & \alpha.(p + q)  &\sim & \alpha.p + \alpha.q & \quad &\;  \mbox{(distributivity)}\\
\qquad & \epsilon.r  &\sim & r &  & \; \mbox{($\epsilon$-absorption)}
\end{array}$
\proof The proof follows directly from Proposition \ref{prop-obvious}. The law
$\alpha.\nil  \sim  \nil$ holds because $L(\alpha.\nil) = \{\alpha w \mid w \in L(\nil)\} =
\{\alpha w \mid w \in \emptyset\} = \emptyset = L(\nil)$. The law $\alpha.(p + q)  \sim \alpha.p + \alpha.q$
holds because $L(\alpha.(p + q)) = \{\alpha w \mid w \in L(p+q)\} =$ $ \{\alpha w \mid w \in L(p) \cup L(q)\} =
\{\alpha w \mid w \in L(p)\} \cup \{\alpha w \mid w \in L(q)\} $ $= L(\alpha.p) \cup L(\alpha.q) =  L(\alpha.p + \alpha.q)$. The law 
$\epsilon.r  \sim r$ holds because $L(\epsilon.r) = \{\epsilon w \mid w \in L(r)\} = \{w \mid w \in L(r)\} = L(r)$.
\fine
 \end{proposition}

It is easy to see that none of the laws for prefixing hold for bisimilarity, and so not even for isomorphism equivalence.

We now focus on the properties of constants. First, the unfolding law (which holds also for bisimilarity), then the excision law 
(useful to remove loops of $\epsilon$ transitions)
and finally, in the next subsection, the folding law, whose proof relies on some auxiliary technical machinery (fixed point theorems on a
complete lattice, see, e.g., \cite{DP02,San10}).

\begin{proposition}\label{rec-law-unf}{\bf (Unfolding)}
 For each $p \in  \mathcal{P}_{SFM1}^{grd}$ and each $C \in \cons$, 
if $C \eqdef p$ then $C   \sim  p$.
\proof Trivial by construction in Table \ref{den-nfa-sfm1}.
\fine
 \end{proposition}

The unfolding law also holds for bisimilarity \cite{Mil89,GV15,Gor19}, but it does not hold for isomorphism equivalence.
For instance, consider $C \eqdef a.a.C + \1$: it is easy to see that the NFA for $C$ has two states, while the NFA for $a.a.C +\1$
has three states.

\begin{remark}\label{eps-rem}
The $\epsilon$-absorption law in Proposition \ref{pref-alg-prop} is better split in two separate laws:
\begin{itemize}
\item[$(i)$] $\epsilon.p \sim p\; $ if $\; p \neq C$, and
\item[$(ii)$] $\epsilon.C \sim p\; $ if $\; C \eqdef p$,
\end{itemize}

\noindent
where the second law is derivable from the first one and the unfolding law. In fact, 
if $C \eqdef p$, then by unfolding we have
$C \sim p$; then, by congruence, $\epsilon.C \sim \epsilon.p$ and then, by the first law, $\epsilon.p \sim p$, so that the equality
$\epsilon.C \sim p$ follow by transitivity. 

The usefulness of the first law $(i)$ instead of the more general $\epsilon$-absorption
law in Proposition \ref{pref-alg-prop}
 is that it preserves guardedness; for instance, 
if $C \eqdef \epsilon.D + \1$ and $D \eqdef a.\1$, then by using (unfolding and) the more general law we 
can derive $C \sim D +\1$, where the r.h.s. is not a guarded term; on the contrary, by using (recursion congruence and) $(i)$, 
we can derive $C \sim a.\1 + \1$, where the r.h.s. is a guarded term.
\fine
\end{remark}

\begin{proposition}\label{rec-law-exc}{\bf (Excision)}
For each $p \in  \mathcal{P}_{SFM1}^{grd}$, and each $C, D \in \cons$,
if $C \eqdef  (\epsilon.x + p)\{C/x\}$ and $D \eqdef p\{D/x\}$, then $C \sim D$.
\proof Trivial, as the self-loop $\epsilon$-labeled transition does not contribute to the generation of any word.
\fine
\end{proposition}

Of course, the excision law does not hold for bisimilarity, and so not even for isomorphism equivalence. 

%
\subsection{The Folding Law}\label{folds-sec}
%

In order to prove the folding law, we first observe that $(\mathcal{P}(A^*), \subseteq)$ is a {\em complete} lattice, i.e., with the property that
all the subsets of $\mathcal{P}(A^*)$ ( e.g., $\{L_i\}_{i\in I}$) have joins (i.e., $\bigcup_{i \in I} L_i$) and meets
(i.e., $\bigcap_{i \in I} L_i$), whose top element $\top$ is $A^*$ and bottom element $\bot$ is $\emptyset$.

At a first approximation, the folding law states that if $C \eqdef p\{C/x\}$ and $q \sim p\{q/x\}$, then $ C  \sim q$.
Of course, we can easily realize that $L(C) = L_p^{\downarrow} \cup (L_p^x \cdot L(C))$ as well as
 $L(q) = L_p^{\downarrow} \cup (L_p^x \cdot L(q))$. Therefore, both $L(C)$ and $L(q)$ are a fixed point of the
 mapping $T(W) = L_p^{\downarrow} \cup (L_p^x \cdot W)$. We want to show that, under one mild condition, 
 there exists exactly one fixed point of
 $T$, so that $L(C) = L(q)$ and so 
 $C \sim q$, as required. 
 
If the mapping $T$ is {\em continuous} (i.e.,
for all increasing sequences $L_0 \subseteq L_1 \subseteq L_2 \ldots$ we have that $T(\bigcup_{i} L_i) = \bigcup_i T(L_i)$),
then the {\em least fixed point} {\tt lfp}$(T) = \bigcup_{n \geq 0} T^n(\emptyset)$, where $T^0(L) = L$ and $T^{n+1}(L) = T(T^n(L))$.
Indeed, $T$ is continuous because $T(\bigcup_{i} L_i) = L_p^{\downarrow} \cup (L_p^x \cdot \bigcup_{i} L_i)$
$= L_p^{\downarrow} \cup \bigcup_{i} (L_p^x \cdot L_i) = $ $\bigcup_{i} ( L_p^{\downarrow} \cup (L_p^x \cdot L_i)) = \bigcup_i T(L_i)$.
The least fixed point {\tt lfp}$(T) = \bigcup_{n \geq 0} T^n(\emptyset)$ can be computed as follows:\\

$\begin{array}{rcl}
T^0(\emptyset) & = & \emptyset\\
T^1(\emptyset) & = & L_p^{\downarrow}\\
T^2(\emptyset) & = & L_p^{\downarrow} \cup (L_p^x \cdot L_p^{\downarrow})\\
T^3(\emptyset) & = & L_p^{\downarrow} \cup (L_p^x \cdot L_p^{\downarrow}) \cup (L_p^x \cdot (L_p^{\downarrow} \cup (L_p^x \cdot L_p^{\downarrow})) = L_p^{\downarrow} \cup (L_p^x \cdot L_p^{\downarrow}) \cup L_p^x \cdot (L_p^x \cdot L_p^{\downarrow})\\
\ldots\\
\end{array}$

\noindent whose union is the language {\tt lfp}$(T) = (L_p^x)^* \cdot L_p^{\downarrow}$.
 
Dually, if $T$ is {\em co-continuous} (i.e., 
for all decreasing sequences $L_0 \supseteq L_1 \supseteq L_2 \ldots$ we have that $T(\bigcap_{i} L_i) = \bigcap_i T(L_i)$),
then the {\em greatest fixed point} {\tt gfp}$(T) = \bigcap_{n \geq 0} T^n(A^*)$.
Indeed, $T$ is co-continuous because $T(\bigcap_{i} L_i) = L_p^{\downarrow} \cup (L_p^x \cdot \bigcap_{i} L_i)$
$= L_p^{\downarrow} \cup \bigcap_{i} (L_p^x \cdot L_i) = $ $\bigcap_{i} ( L_p^{\downarrow} \cup (L_p^x \cdot L_i)) = \bigcap_i T(L_i)$.
The greatest fixed point {\tt gfp}$(T) = \bigcap_{n \geq 0} T^n(A^*)$ can be computed as follows:\\

$\begin{array}{rcl}
T^0(A^*) & = & A^*\\
T^1(A^*) & = & L_p^{\downarrow} \cup (L_p^x \cdot A^*)\\
T^2(A^*) & = & L_p^{\downarrow} \cup (L_p^x \cdot (L_p^{\downarrow} \cup (L_p^x \cdot A^*)) = L_p^{\downarrow} \cup (L_p^x \cdot L_p^{\downarrow}) \cup L_p^x \cdot (L_p^x \cdot A^*)\\
T^3(A^*) & = & L_p^{\downarrow} \cup (L_p^x \cdot 
(L_p^{\downarrow} \cup (L_p^x \cdot L_p^{\downarrow}) \cup L_p^x \cdot (L_p^x \cdot A^*)) = $ $\\
& & L_p^{\downarrow} \cup (L_p^x \cdot L_p^{\downarrow}) \cup L_p^x \cdot (L_p^x \cdot L_p^{\downarrow}) \cup 
L_p^x \cdot (L_p^x \cdot (L_p^x \cdot A^*))\\
\ldots\\
\end{array}$

\noindent whose intersection is the language {\tt gfp}$(T) = (L_p^x)^* \cdot L_p^{\downarrow}$, unless $L_p^x$ does contain
 the empty word $\epsilon$, because in such a case the application of $T$ does not restrict the newly computed language,
 so that $T^n(A^*) = A^*$ for all $n \geq 0$, and so the greatest fixed point of $T$ is the top element
 $A^*$ of the complete lattice.
 We can characterize this condition syntactically as follows.

\begin{definition}{\bf (Observationally guarded)}\label{og-def}
A (possibly open) SFM1 process $p$ is {\em observationally guarded} if $og(p)$ holds, 
where $og(-)$ is defined as the least predicate generated by the axioms and the rules in Table \ref{og-tab}.
\fine
\end{definition}

\begin{table}[t]
{\renewcommand{\arraystretch}{1.8}

\hrulefill\\[-.9cm]

\begin{center}\fontsize{9pt}{0.08pt}

$\begin{array}{llllllllllll}
\bigfrac{}{og(\nil)} &  \bigfrac{}{og(\1)} &  \bigfrac{a \in A}{og(a.p)}  &\bigfrac{og(p)}{og(\epsilon.p)} 
&  \bigfrac{og(p) \quad C \eqdef p\{C/x\}}{og(C)} &  \bigfrac{og(p) \quad og(q)}{og(p + q)}\\
\end{array}$

\hrulefill\\
\end{center}
\caption{Observationally guarded predicate}\label{og-tab}}
\end{table}

Note that $C \eqdef p\{C/x\}$ is observationally guarded if and only if $C \NDeriv{\epsilon} C$, or equivalently, if and 
only if $p \NDeriv{\epsilon} x$, i.e., if and only if $\epsilon \not \in L_p^x$. 


\begin{proposition}\label{rec-law-fold}{\bf (Folding)}
 For each $p \in  \mathcal{P}_{SFM1}^{grd}$ (open on $x$), and each $C \in \cons$, 
 if $C \eqdef p\{C/x\}$, $og(p)$ and $q \sim p\{q/x\}$, then $ C  \sim q$.
 \proof If $p$ is observationally guarded, the least fixed point and the greatest fixed point of the function
 $T(W) = L_p^{\downarrow} \cup (L_p^x \cdot W)$ do coincide, so that the fixed point is unique.
 Since both $L(C)$ and $L(q)$ are fixed points for $T$, then $L(C) = L(q)$.
 \fine
 \end{proposition}

Observe that the requirement that $p$ is observationally guarded is crucial for the 
correctness of the folding property.
For instance, consider $p = \epsilon.x + a.\1$, which is not observationally guarded. 
Then, $C \eqdef \epsilon.C + a.\1$ defines a process that can recognize only $a$, so it is 
language equivalent to $a.\1$. 
However, $q = a.\1 + b.\1$ is such that $q \sim \epsilon.q + a.\1$,
but $C \not \sim q$.


\begin{remark}\label{strong-fold-rem}{\bf (Folding Law for Bisimulation Equivalence)}
It is possible to prove, with a standard bisimulation-based proof technique \cite{Mil84,Gor19}, that the more concrete 
bisimulation equivalence $\simeq$  satisfies a stronger
folding law: For each $p \in  \mathcal{P}_{SFM1}^{grd}$ (open on $x$), and each $C \in \cons$, 
 if $C \eqdef p\{C/x\}$ and $q \simeq p\{q/x\}$, then $ C  \simeq q$.
\fine
\end{remark}

\section{Equational Deduction with Process Constants}\label{eq-ded-sec}

In the following, we show that it is possible to prove syntactically, i.e., by means of an equational deductive proof, when two SFM1 processes recognize the same language. 
First, in this section we describe how equational deduction is implemented in a process algebra with process constants;
 then in the next section we provide a sound and complete axiomatization of language equivalence for observationally guarded SFM1 processes, composed of 7 axioms and 2 conditional axioms; 
 finally, we show that each observationally unguarded process can be equated to an observationally guarded one with the help of one additional conditional axiom. Hence, the finite axiomatization of language equivalence over SFM1 is composed of 7 axioms and 3 conditional axioms, only.
 Let us start with an equational deduction primer.\\

Given a (one-sorted) {\em signature} $\Sigma$ (i.e., a finite set of constants, a finite set of process constants 
and a finite set of function symbols with their arities),
we can define the set of terms over $\Sigma$ and over a finite set of variables $V$ (notation $\mathcal{T}(\Sigma, V)$)
as the smallest set that satisfies the following:

\begin{itemize}
	\item[$(i)$]  each variable $x$ in $V$ is a term in $\mathcal{T}(\Sigma, V)$;
	\item[$(ii)$]   each constant $c$ in $\Sigma$ is a term of $\mathcal{T}(\Sigma, V)$;
	\item[$(iii)$] each process constant $C$ in $ \Sigma $ is a term of $\mathcal{T}(\Sigma, V)$;
	moreover, each constant $C$ must be equipped with a definition, e.g., $C \eqdef t$, 
	with $t$ a term in $\mathcal{T}(\Sigma, V)$;
	\item[$(iv)$]  if $f$ is a $n$-ary function symbol and $t_1, \ldots, t_n$ are terms in $\mathcal{T}(\Sigma, V)$,
	then also $f(t_1, \ldots, t_n)$ is a term in $\mathcal{T}(\Sigma, V)$.
\end{itemize}

Moreover, we have that $\mathcal{T}(\Sigma) $ denotes the set of {\em closed} (or {\em ground}) terms, 
where a term $t \in \mathcal{T}(\Sigma) $ when:

\begin{itemize}
	\item there are no variable occurrences in $t$
	\item $ \forall C \in \const{t} $, if $ C \eqdef p$, then $ p \in \mathcal{T}(\Sigma)$.
\end{itemize} 
 The definition is apparently recursive, but this is not a problem, because we can define it keeping track of 
 already encountered process constants, like we did in the denotational semantics of SFM1.
 
 A {\em substitution} $\rho$ is a mapping from $V$ to $\mathcal{T}(\Sigma, V)$. For any term $t$, by $t[\rho]$ we denote
the term obtained as follows:

\begin{itemize}
	\item[$(i)$] for each variable $x$ in $V$, $x[\rho] = \rho(x)$,
	\item[$(ii)$] for each constant $c$ in $\Sigma$, $c[\rho] = c$,
	\item[$(iii)$] for each process constant $C$ in $ \Sigma $ such that $C \eqdef t $, $ C[\rho] =C $ with $ C \eqdef t[\rho] $
	\item[$(iv)$] for any $n$-ary function symbol $f$ and terms  $t_1, \ldots, t_n$, $f(t_1, \ldots, t_n)[\rho]$
	is the term  $f(t_1[\rho], \ldots, t_n[\rho])$.
\end{itemize}

A substitution $\rho$ is {\em closed} if $\rho(x) \in \mathcal{T}(\Sigma)$ for each variable $x \in V$. Hence, the application of a closed substitution to an open term yields a closed (or ground) term. \\

\begin{table}[t]
	{{\renewcommand{\arraystretch}{0.9}
			\hrulefill\\[-.9cm]
			\begin{center}\fontsize{8pt}{-.1pt}
				$\begin{array}{llllll}
				1. &\; \;  \mbox{Reflexivity} &\; \;  \bigfrac{}{ t \, = \, t}\\
				2. &\; \;  \mbox{Symmetry} &\; \;  \bigfrac{t_1 \, = \, t_2}{ t_2 \, = \, t_1}\\
				3. &\; \;  \mbox{Transitivity} &\; \;  \bigfrac{t_1 \, = \, t_2 \; \; \; t_2 \, = \, t_3}{ t_1 \, = \, t_3}\\
				4.1 &\; \;  \mbox{Substitutivity} &\; \;  \bigfrac{t_i \, = \, t'_i}{f(t_1, \ldots  t_i,  \ldots t_k)  \, = \, f(t_1, \ldots  t'_i, 
				\ldots t_k)} 
				& \mbox{for any $f$ and $1 \leq i \leq n$}\\
				4.2 &\; \;  \mbox{Recursion} &\; \;  \bigfrac{t_1 = t_2 \wedge C \eqdef t_1\{C/x\} \wedge D \eqdef t_2\{D/x\}}{C = D} & \\
				5. &\; \;  \mbox{Instantiation} &\; \;  \bigfrac{t_1 \,=\, t_2}{ t_1[\rho] \,=\, t_2[\rho]} & \mbox{for any substitution $\rho$}\\ 
				6. &\; \;  \mbox{Axioms} &\; \;  \bigfrac{}{ t_1 \, = \, t_2} & \mbox{for all axioms $t_1= t_2$ in $E$}\\
				\end{array}$
				
				\hrulefill
		\end{center}}
	}
	\caption{Rules of equational deduction}\label{eq-ded}
\end{table}

\begin{remark}\label{sig-sfm-rem}
The (one-sorted) {\em signature} $\Sigma$ for our language {\em open} SFM1 is as follows:

\begin{itemize}
	\item variables: $ x \in Var $ (where $ Var $ is finite);
	\item two constants: $\1$ and $\nil$;
	\item process constants: $ C \in \cons $ (where $ \cons $ is finite);
	\item operators:
	\begin{itemize}
		\item $ \alpha.\_ $ (a finite family of prefixing operators, one for each $\alpha \in A \cup \{\epsilon\} $), 
		\item $\_+\_ $ (one binary operator of alternative composition).
	\end{itemize}
\end{itemize}

This means that, in this section, the syntactic definition of open SFM1 is given with only one syntactic 
category in the following way:\\

$\begin{array}{lccccccccccccccl}
p &  ::= &  \1 & | &  \nil & | & \alpha.p & | &  p+p & | & C & | & x\\
\end{array}$\\

\noindent so that it is possible to write terms, such as $x + (y + z)$, which cannot be described in the syntax
of open SFM1 of Section \ref{cong-sec}.
\fine
\end{remark}

An {\em equational theory} is a tuple $(\Sigma, E)$, where $\Sigma$ is a signature and $E$ is a set of equations
of the form $t_1 \, = \; t_2$, where $t_1$ and $t_2$ are terms in $\mathcal{T}(\Sigma, V)$. The equations in $E$ are
usually called {\em axioms} and the equational theory $(\Sigma, E)$ is often called an {\em axiomatization}.

For any equational theory $(\Sigma, E)$, there is a standard set of proof rules for deriving equalities 
on $\mathcal{T}(\Sigma, V)$. This set of proof rules forms an 
{\em equational deductive proof system} $D(E)$, composed of the rules in Table \ref{eq-ded}.

A {\em proof} is a finite sequence of equalities $t_1 \, = \, t'_1, \; t_2 \, = \, t'_2,  \ldots t_k \, = \, t'_k$ such that each $t_i \, = \, t'_i$
is either an axiom (rule $1$ or $6$) or can be derived by using one of the rules $2 - 5$ with premises some of the 
previous equalities $t_1 \, = \, t'_1, \; t_2 \, = \, t'_2,  \ldots t_{i - 1} \, = \, t'_{i - 1}$.
This linear representation of a proof can be more conveniently expressed in the form of a tree (proof tree).

We use the notation $E \vdash t_1 \, = \, t_2$ if there exists a proof with $ t_1 \, = \, t_2$ as its last equality, using only
the axioms in $E$ (plus the axiom in rule $1$). This directly determines a congruence over the set of terms: 
$t_1 =_E t_2$ if and only if $E \vdash t_1 \, = \, t_2$. Indeed, note that rules $1-3$ ensure that the relation $ =_E$ that $D(E)$ is 
inducing on $\mathcal{T}(\Sigma, V)$ is an equivalence relation;
similarly, rules $4.1$ and $4.2$ ensures substitutivity of equals for equals in any context, i.e., that the relation 
$ =_E$ that $D(E)$ is inducing is a congruence. 

Let $S \subseteq \mathcal{T}(\Sigma)$ be the set of terms of interest (we may think that $S$ is the set $\mathcal{P}_{SFM1}$ of 
SFM1 processes, which is included in the larger set of all the closed terms of the syntax of open SFM1 
in Remark \ref{sig-sfm-rem}). 

Let $R \subseteq  S \times  S$ be a relation on closed (or ground) terms. 
The equational deductive proof system $D(E)$ for the equational theory $(\Sigma, E)$ is (ground) {\em sound} w.r.t. $R$ if 
\[\forall (t_1, t_2) \in S \times S, \quad E \vdash t_1 \, = \, t_2 \; \; \mbox{ implies } \; \; (t_1, t_2) \in R.\]
$D(E)$ is (ground) {\em complete} w.r.t. $R$ if
\[(t_1, t_2) \in R  \; \; \mbox{ implies } \; \; E \vdash t_1 \, = \, t_2.\]
In other words, a sound and complete axiomatization $(\Sigma, E)$ of a congruence $R$ on $ S \subseteq \mathcal{T}(\Sigma)$ 
is such that the congruence $t_1 =_E t_2$ induced by $D(E)$ on $S$ is exactly the same congruence defined by $R$.

%
\section{Axiomatization}\label{axiom-sec}
%
 
 In this section we present a sound and complete, finite axiomatization of language equivalence over SFM1. 
As hinted in the previous section (cf. Remark \ref{sig-sfm-rem}), for simplicity's sake, the syntactic definition of open SFM1 is given with only 
one syntactic category, but 
each ground instantiation of an axiom must respect the syntactic definition of SFM1 given in 
Section \ref{cong-sec}; this means that we can write the axiom $x + (y + z) = (x + y) + z$, but 
it is invalid to instantiate it to $C + (a.\1 + b.\1) = (C + a.\1) + b.\1$ because these are not legal 
SFM1 processes (the constant $C$ cannot be used as a summand).

\subsection{The Axioms}\label{set-axiom-sec}

The set of axioms is outlined in Table \ref{axiom-sfm1-tab}. 
The axioms {\bf A1-A4} are the usual axioms for choice \cite{Kleene,Salomaa,Kozen,Mil84}. 
The axiom schemata {\bf T1-T3} are those for prefixing, that are finitely many
as $A$ is finite. Note that {\bf T3} has the side condition $x \neq C$, meaning that it can be used only if $x$ is not 
instantiated to a process constant: this is necessary to preserve constant guardedness (see Remark \ref{eps-rem}).
The conditional axioms {\bf R1-R3} are about process constants and are similar to those in \cite{Mil84,Mil89a,Gor19,Gor20ic}. 
Note that these conditional axioms are actually a finite collection of axioms, one for each constant 
definition: since the set $\cons$ of process constants is finite, the instances of {\bf R1-R3} are finitely many as well. 

The folding axiom {\bf R2} has some similarities with this `iterative' conditional axiom, due to Arden \cite{Arden} and 
Salomaa \cite{Salomaa}:

$x = (y \cdot x) + z \; \Rightarrow \; x = y^*z$,

\noindent
 provided that $y$ does not have the {\em empty word property} (ewp, for short), meaning that $\epsilon \not\in L(y)$.
 Kozen \cite{Kozen} considers this conditional axiom not algebraic, in the sense that it is not preserved by substitution.
The side condition $og(p)$ (being equivalent to $\epsilon \not \in L^x_p$) of {\bf R2} is essentially the negation of the ewp property.
However, the conditional axiom {\bf R2} is algebraic because the term $p$ is a concrete one (not a variable), depending on the actual 
 definition of the constant $C$.
The excision axiom {\bf R3} has some similarities
with one law proposed by Salomaa in \cite{Salomaa}, namely  $(x + \1)^* = x^*$.

Besides the folding axiom {\bf R2}, we consider also a stronger variant, called {\bf R2$'$}, where the side condition $og(p)$ is omitted:\\

{\bf R2$'$}   (B-Folding) \quad  if $C \eqdef p\{C/x\}  \; \wedge q = p\{q/x\}$ then $C = q$\\

\noindent which holds for bisimilarity \cite{Mil84,Gor19}.

We call $B$ the set of axioms $\{${\bf A1, A2, A3, A4, R1, R2$'$}$\}$ that are sound and complete for bisimilarity \cite{Mil84,Gor19,Gor20ic}, while we call
$W_g$ the set of axioms $\{${\bf A1, A2, A3, A4, T1, T2, T3, R1, R2}$\}$ that we will
prove to be a sound and complete axiomatization of language equivalence
for observationally guarded SFM1 processes, while $W = W_g \cup \{${\bf R3}$\}$ is the full axiomatization of language equivalence
over the whole of SFM1.
By the notation $W \vdash p = q$ we mean that there exists an equational deduction proof 
of the equality $p = q$, by using the axioms in $W$.

\begin{table}[t]
{\renewcommand{\arraystretch}{1.2}
\hrulefill\\[-.7cm]
			\begin{center}\fontsize{8pt}{0.1pt}

$\begin{array}{llrcll}
{\bf A1} &\; \;  \mbox{Associativity} &\; \;  x + (y + z) & = & (x + y) + z &\\
{\bf A2} &\; \;  \mbox{Commutativity} &\; \;  x + y & = & y + x& \\
{\bf A3} &\; \;  \mbox{Identity} &\; \;  x + \nil & = & x & \\
{\bf A4} &\; \;  \mbox{Idempotence} &\; \;  x + x & = & x & \\
\end{array}$

\hrulefill

$\begin{array}{llrcll}
{\bf T1} &\; \;  \mbox{Annihilation} &\; \;  \alpha.\nil & = & \nil & \\
{\bf T2} &\; \;  \mbox{Distributivity} &\; \;  \alpha.(x + y) & = & \alpha.x + \alpha.y& \\
{\bf T3} &\; \;  \mbox{$\epsilon$-absorption} &\; \;  \epsilon.x & = & x &  \quad  \mbox{ if $x \neq C$}\\
\end{array}$

\hrulefill

$\begin{array}{llrcllll}
{\bf R1} &\; \mbox{Unfolding} &\; \; \mbox{if $C \eqdef p \;$} & \mbox{then} &
\mbox{$C = p$} &\\
{\bf R2} & \;  \mbox{Folding} &  \mbox{if $C \eqdef p\{C/x\}  \; \wedge \; og(p) \; \wedge \;
q = p\{q/x\}$} & \mbox{then} & \mbox{$C = q$} & \\
{\bf R3} &\;  \mbox{Excision} &  \mbox{if $C \eqdef  (\epsilon.x + p)\{C/x\} \wedge D \eqdef p\{D/x\}$} & \mbox{then} &
C =  D &\\
\end{array}$

\hrulefill

\end{center}
}
\caption{Axioms for language equivalence}\label{axiom-sfm1-tab}
\end{table}

\begin{theorem}{\bf (Soundness)}\label{sound-th-sfm1}
For every $p, q \in  \mathcal{P}_{SFM1}$, if $W \vdash p = q$, then $p \sim q$.
\proof The proof is by induction on the proof of $W \vdash p = q$. The 
thesis follows by observing that all the axioms in $W$ are sound  (e.g., {\bf T1-T3} by Proposition \ref{pref-alg-prop}) 
and that $ \sim$ is a congruence.
\fine
\end{theorem} 

\subsection{Normal Forms, Unique Solutions and $\epsilon$-free Normal Forms}\label{det-nf-sec}

An SFM1 process $p$ is a {\em normal form} if the predicate $nf(p)$ holds. This predicate stands for 
$nf(p, \emptyset)$, whose inductive definition is displayed in Table \ref{nf1-tab}. 
Examples of terms which are not in normal form are  $a.b.\1$ and $C \eqdef a.b.C + \1$.

\begin{table}[t]
{\renewcommand{\arraystretch}{1.6}
\hrulefill\\[-.6cm]
			\begin{center}\fontsize{8pt}{0.1pt}

$\begin{array}{cccccc}
\bigfrac{}{nf(\1, I)} & \quad & \bigfrac{}{nf(\nil, I)} & \quad    & 
\bigfrac{nf(p, I \cup \{C\}) \quad C \eqdef p \quad C \not \in I}{nf(C, I)}\\

\bigfrac{nf(C, I)}{nf(\alpha.C, I)}
& \quad & \bigfrac{nf(p, I) \quad nf(q, I)}{nf(p + q, I)} & \quad & \bigfrac{C \in I}{nf(C, I)} \\
\end{array}$

\hrulefill\\
\end{center}}
\caption{Normal form predicate}\label{nf1-tab}
\end{table}

Note that
if $C$ is a normal form, then its body (ignoring all the possible summands $\nil$ or duplicated summands 
$\1$, that can be absorbed via axioms {\bf A3-A4})
is of the form $\sum_{i = 1}^{n} \alpha_i.C_i $ or $\sum_{i = 1}^{n} \alpha_i.C_i + \1$,
(assuming that $\sum_{i = 1}^{n} \alpha_i.C_i$ is $\nil$ if $n = 0$) where, in turn, each $C_i$ is a normal form.
As a matter of fact, according to the construction in the proof of Theorem \ref{representability-nfa}, the processes representing 
reduced NFAs are in normal form.

Now we show that, for each SFM1 process $p$, there exists a normal form $q$ such that $B \vdash p = q$. Since bisimilarity is
finer than language equivalence, this ensures that the normal form $q$ is language equivalent to $p$. 

\begin{proposition}\label{nf-prop}{\bf (Reduction to normal form)}
Given an SFM1 process $p$, there exists a normal form $q$ such that $B \vdash p = q$.

\proof The proof is by induction on the structure of $p$, with the proviso to use a set $I$ of already scanned constants, 
in order to avoid looping on recursively defined constants, where $I$ is initially empty. 
We prove that for $(p, I)$ there exists a term $q$ such that $nf(q,I)$ holds
and $(B, I) \vdash p = q$, where this means that the equality $p = q$ can be derived by the axioms in $B$ when
each constant $C \in I$ is assumed to be equated to itself only. The thesis then follows by considering $(p, \emptyset)$.

The first base case is $(\1, I)$; in such a case, $q = \1$, because $nf(\1, I)$ holds, and the thesis 
$(B, I) \vdash \1 = \1$ follows by reflexivity.
The second base case is $(\nil, I)$ and it is analogous to the previous one.

Case $(\alpha.p, I)$: by induction, $(p, I)$ has an associated normal form $nf(q, I)$ such that $(B, I) \vdash p = q$;
hence, $(B, I) \vdash \alpha.p = \alpha.q$ by substitutivity. If $q$ is a constant, then $\alpha.q$ is already a 
normal form. Otherwise,
take a new constant $C \eqdef q$, so that $nf(C, I)$ holds because $nf(q, I \cup \{C\})$ holds
(note that $C$ does not occur in $q$, so that this is the same as stating $nf(q, I)$, which holds by induction).
The required normal form is $\alpha.C$. Indeed, $nf(\alpha.C, I)$ holds because $nf(C, I)$ holds; moreover,
since $(B, I) \vdash p = q$ by induction and $(B, I) \vdash C = q$ by axiom {\bf R1}, we have that 
$(B, I) \vdash p = C$ by transitivity, so that
$(B, I) \vdash \alpha.p = \alpha.C$ by substitutivity.

Case $(p_1 + p_2, I)$: by induction, we can assume that for $(p_i, I)$ there exists a normal form $nf(q_i, I)$
such that $(B, I) \vdash p_i = q_i$ for $i = 1, 2$. Then, $nf(q_1 + q_2, I)$ holds and 
$(B, I) \vdash p_1 + p_2 = q_1 + q_2$ by substitutivity.

Case $(C, I)$, where $C \eqdef r\{C/x\}$. If $C \in I$, then we can stop induction, by returning $C$ itself: $nf(C, I)$ holds
and $(B, I) \vdash C = C$ by reflexivity.
Otherwise (i.e., if $C \not\in I$), by induction on $(r\{C/x\}, I \cup \{C\})$, we can assume that there exists 
a normal form $q\{C/x\}$ such that $nf(q\{C/x\}, I \cup \{C\})$ holds and
$(B, I \cup \{C\}) \vdash r\{C/x\} = q\{C/x\}$. Note that, by construction, if $(B, I \cup \{C\}) \vdash r\{C/x\} = q\{C/x\}$,
then also $(B, I) \vdash r\{C/x\} = q\{C/x\}$. Hence, as $(B, I) \vdash C =  r\{C/x\}$ by axiom {\bf R1}, it follows that
$(B, I) \vdash C = q\{C/x\}$ by transitivity. 
Then, we take a new constant $D \eqdef q\{D/x\}$ such that  $nf(D, I)$ because $nf(q\{D/x\}, I \cup \{D\})$ holds.
Hence, $(B, I) \vdash C = D$ by axiom {\bf R2$'$}, where $D$ is
a normal form.
\fine
\end{proposition}

\begin{corollary}\label{nf-cor}{\bf (Reduction to observationally guarded normal form)}
Given an observationally guarded SFM1 process $p$, there exists an observationally guarded normal form $q$ 
such that $W_g \vdash p = q$.

\proof The proof is very similar to that of the previous proposition.
We prove that for $(p, I)$ such that $og(p)$ there exists a term $q$ such that $nf(q,I)$ and $og(q)$ hold
and $(W_g, I) \vdash p = q$, where this means that the equality $p = q$ can be derived by the axioms in $W_g$ when
each constant $C \in I$ is assumed to be equated to itself. The thesis then follows by considering $(p, \emptyset)$.

All the cases are trivial adaptation of the corresponding above, except the one for recursion.
 
Case $(C, I)$, where $C \eqdef r\{C/x\}$ and $og(C)$ holds. If $C \in I$, then we can stop induction, 
by returning $C$ itself: $nf(C, I)$ holds
and $(W_g, I) \vdash C = C$ by reflexivity.
Otherwise (i.e., if $C \not\in I$), by induction on $(r\{C/x\}, I \cup \{C\})$ with $og(r)$, we can assume that there exists 
a normal form $q\{C/x\}$ such that $nf(q\{C/x\}, I \cup \{C\})$ and $og(q)$ hold and, moreover,
$(W_g, I \cup \{C\}) \vdash r\{C/x\} = q\{C/x\}$. Note that, by construction, if $(W_g, I \cup \{C\}) \vdash r\{C/x\} = q\{C/x\}$,
then also $(W_g, I) \vdash r\{C/x\} = q\{C/x\}$. Hence, as $(W_g, I) \vdash C =  r\{C/x\}$ by axiom {\bf R1}, it follows that
$(W_g, I) \vdash C = q\{C/x\}$ by transitivity. 
Then, we take a new constant $D \eqdef q\{D/x\}$ such that  $nf(D, I)$ because $nf(q\{D/x\}, I \cup \{D\})$ holds, and
$og(D)$ holds because $og(q)$ holds.
Hence, $(W_g, I) \vdash C = D$ by axiom {\bf R2}, where $D$ is
an observationally guarded normal form.
\fine
\end{corollary}

\begin{remark}{\bf (Normal forms as systems of equations)}\label{rem-normal}
As a matter of fact, we can restrict our attention only to 
normal forms defined by constants: if $p$ is a normal form and 
$p$ is not a constant, then take a new constant $D \eqdef p$, which is a normal form such that $B \vdash D = p$
by axiom {\bf R1}. For this reason, 
in the following we restrict our attention to
normal forms that can be defined by means of a {\em system of equations}:  a set 
$\widetilde{C} = \{C_1, C_2, \ldots, C_n\}$ of defined constants in normal form,
such that $\const{C_1} = \widetilde{C}$ and $\const{C_i} \subseteq \widetilde{C}$ for $i = 1, \ldots, n$.
The equations of the system $E(\widetilde{C})$ 
are of the following form (where function
$f(i,j)$ returns a value $k$ such that $1 \leq k \leq n$):\\

$\begin{array}{lcllcllcl}
C_1 & \eqdef &  \sum_{i = 1}^{m(1)} \alpha_{1i}.C_{f(1,i)} \{+\1\}\\
C_2 & \eqdef &  \sum_{i = 1}^{m(2)} \alpha_{2i}.C_{f(2,i)} \{+\1\}\\
\ldots\\
C_n & \eqdef & \sum_{i = 1}^{m(n)} \alpha_{ni}.C_{f(n,i)} \{+\1\}\\
\end{array}$\\

\noindent
where by the notation $\{+\1\}$ we mean that the additional summand $\1$ is optional, and where
we assume $C_i \eqdef \nil$ in case $m(i) = 0$. We sometimes use the notation $body(C_h)$ to denote
the sumform $ \sum_{i = 1}^{m(h)} \alpha_{hi}.C_{f(h,i)} \{+\1\}$.
\fine
\end{remark}

\begin{remark}\label{rem-obs-g}{\bf (Observationally guarded system of equations)}
We say that a system of equations $E(\widetilde{C})$, not necessarily in normal form
like the one above, is 
{\em observationally guarded} if, for each $C_i \in \widetilde{C}$, there is no silent cycle $C_i \Deriv{\epsilon} C_i$. 
This definition is coherent with Definition \ref{og-def}.
For instance, consider the following example:

$\begin{array}{lcllcllcl}
C_1 & \eqdef & a.C_2 + \epsilon.a.C_1\\
C_2 & \eqdef &  b.C_2 + \epsilon.C_1\\
\end{array}$

\noindent
Of course, $C_i \NDeriv{\epsilon} C_i$ for $i = 1, 2$, so that the system is observationally guarded. And indeed
we can prove that $og(C_1)$ and $og(C_2)$ hold.
\fine
\end{remark}

The next theorem shows that every observationally guarded system of equations 
(not necessarily in normal form) has a unique solution up to provably equality.

\begin{theorem}{\bf (Unique solution for $W_g$)}\label{unique-wg-th}
Let $\widetilde{X} = (x_1, x_2, \ldots, x_n)$ be a tuple of variables and let 
$\widetilde{p} = (p_1, p_2, \ldots, p_n)$ be a tuple of
open guarded SFM1 terms, using the variables in $\widetilde{X}$. 
Let $\widetilde{C} = (C_1, C_2, \ldots, C_n)$ be a tuple of constants (not occurring in $\widetilde{p}$) 
such that the system of equations
 
 $\begin{array}{lcllcllcl}
C_1 & \eqdef &  p_1\{\widetilde{C}/\widetilde{X}\} \\
C_2 & \eqdef & p_2\{\widetilde{C}/\widetilde{X}\} \\
\ldots\\
C_n & \eqdef & p_n\{\widetilde{C}/\widetilde{X}\} 
\end{array}$\\

\noindent
is observationally guarded. 
Then, for $i = 1, \ldots, n$, 
$W_g \vdash C_i = p_i\{\widetilde{C}/\widetilde{X}\}$.

Moreover, if the same property holds for $\widetilde{q} = (q_1, q_2, \ldots, q_n)$, i.e.,
for $i = 1, \ldots, n$ $W_g \vdash q_i = p_i\{\widetilde{q}/\widetilde{X}\} $, 
then 
$W_g \vdash C_i = q_i$.

\proof By induction on $n$. 

For $n = 1$, we have $C_1 \eqdef p_1\{C_1/x_1\}$,
and so the result $W_g \vdash C_1 = p_1\{C_1/x_1\}$
follows immediately using axiom {\bf R1}. 
This solution is unique because $p_1$ is observationally guarded 
(cf. Remark \ref{rem-obs-g}): if $W_g \vdash q_1 = p_1\{q_1/x_1\}$, 
by axiom {\bf R2} we get $W_g \vdash C_1 = q_1$.
  
Now assume a tuple $\widetilde{p} = (p_1, p_2, \ldots, p_n)$ and the term $p_{n+1}$,
so that they are all open on $\widetilde{X} = (x_1, x_2, \ldots, x_n)$ and the additional $x_{n+1}$.
Assume, w.l.o.g., that $x_{n+1}$ occurs in $p_{n+1}$. 
First, define 
$C_{n+1} \eqdef p_{n+1}\{C_{n+1}/x_{n+1}\}$, so that $C_{n+1}$ is now open on $\widetilde{X}$.
Therefore, also for $i = 1, \ldots, n$, each $p_{i}\{C_{n+1}/x_{n+1}\}$ is now
open on $\widetilde{X}$.
The resulting observationally guarded
system of equations is:

$\begin{array}{lcllcllcl}
C_1 & \eqdef &  p_1\{C_{n+1}/x_{n+1}\}\{\widetilde{C}/\widetilde{X}\}\\
C_2 & \eqdef & p_2\{C_{n+1}/x_{n+1}\}\{\widetilde{C}/\widetilde{X}\}\\
\ldots\\
C_n & \eqdef & p_n\{C_{n+1}/x_{n+1}\}\{\widetilde{C}/\widetilde{X}\}\\
\end{array}$

\noindent so that now we can use induction on $\widetilde{X}$ and 
$(p_{1}\{C_{n+1}/x_{n+1}\}, \ldots, p_{n}\{C_{n+1}/x_{n+1}\})$, to conclude that 
the tuple $\widetilde{C} = (C_1, C_2, \ldots, C_n)$ of closed constants
is such that for $i = 1, \ldots, n$:

$W_g \vdash C_i = (p_i\{C_{n+1}/x_{n+1}\})\{\widetilde{C}/\widetilde{X}\} = p_{i}\{\widetilde{C}/\widetilde{X}, 
C_{n+1}\{\widetilde{C}/\widetilde{X}\}/x_{n+1}\}$.

\noindent 
Note that above by $C_{n+1}\{\widetilde{C}/\widetilde{X}\}$ we have implicitly closed the definition of 
$C_{n+1}$ as
 
$C_{n+1} \eqdef p_{n+1}\{C_{n+1}/x_{n+1}\}\{\widetilde{C}/\widetilde{X}\} = 
p_{n+1}\{\widetilde{C}/\widetilde{X}\}\{C_{n+1}/x_{n+1}\}$,

\noindent
so that $W_g \vdash C_{n+1} = p_{n+1}\{\widetilde{C}/\widetilde{X}\}\{C_{n+1}/x_{n+1}\}$ 
by axiom {\bf R1}. 

Unicity of the tuple $(\widetilde{C}, C_{n+1})$ can be proved by using axiom {\bf R2}.
Assume to have another solution tuple $(\widetilde{q}, q_{n+1})$. This means that, for $i = 1, \ldots, n+1$,

$W_g \vdash q_i = p_i\{\widetilde{q}/\widetilde{X}, q_{n+1}/x_{n+1}\}$.

\noindent 
By induction, we can assume, for $i = 1, \ldots, n$, that $W_g \vdash C_i = q_i$.

Since $W_g \vdash C_{n+1} = p_{n+1}\{\widetilde{C}/\widetilde{X}\}\{C_{n+1}/x_{n+1}\}$ 
by axiom {\bf R1},
by substitutivity we get $W_g \vdash C_{n+1} = p_{n+1}\{\widetilde{q}/\widetilde{X}\}\{C_{n+1}/x_{n+1}\}$.

Note that $p_{n+1}\{\widetilde{q}/\widetilde{X}\}$ is a term open on $x_{n+1}$ which is observationally guarded.
Let $F$ be a constant defined as follows: $F \eqdef  p_{n+1}\{\widetilde{q}/\widetilde{X}\}\{F/x_{n+1}\}$.
Then, by axiom {\bf R2}, $C_{n+1} = F$.
Hence, since 

$W_g \vdash q_{n+1} = p_{n+1}\{\widetilde{q}/\widetilde{X}\}\{q_{n+1}/x_{n+1}\}$

\noindent
by axiom {\bf R2}, we get $W_g \vdash F = q_{n+1}$; and so  the thesis  
$W_g \vdash C_{n+1} = q_{n+1}$ follows by transitivity.
\fine
\end{theorem}

A system of equations is $\epsilon$-free if there are no occurrences of the prefix $\epsilon$. Formally, an $\epsilon$-free 
normal form must be of the form

$\begin{array}{lcllcllcl}
C_1 & \eqdef &  \sum_{i = 1}^{m(1)} a_{1i}.C_{f(1,i)} \{+\1\}\\
C_2 & \eqdef &  \sum_{i = 1}^{m(2)} a_{2i}.C_{f(2,i)} \{+\1\}\\
\ldots\\
C_n & \eqdef & \sum_{i = 1}^{m(n)} a_{ni}.C_{f(n,i)} \{+\1\}\\
\end{array}$\\

\noindent 
where all the actions $a_{ij} \in A$.

We want to show that an observationally guarded normal form can be equated to an $\epsilon$-free normal form
by using the axioms in $W_g$, in particular {\bf T3}. In fact, assume that

$C_h  \eqdef  \sum_{i = 1}^{m(h)} a_{hi}.C_{f(h,i)}  + \sum_{i = 1}^{m'(h)} \epsilon.C_{g(h,i)} \{+\1\}$

\noindent
and that $C_{g(h,i)} \eqdef p_{g(h,i)}$. Then, by {\bf R1, T3} and substitutivity, we have that

$W_g \vdash C_h  =  \sum_{i = 1}^{m(h)} a_{hi}.C_{f(h,i)}  + \sum_{i = 1}^{m'(h)} p_{g(h,i)} \{+\1\}$.

Of course, in $p_{g(h,i)}$ there may be still occurrences of $\epsilon$, but, since the system of equations is observationally guarded, we 
may proceed as above starting from the constant with no prefix $\epsilon$ and then going backward till reaching the initial constant $C_1$,
as illustrated in the following example.

\begin{example}
Consider the following observationally guarded normal form:

$\begin{array}{rcl}
C_1 & \eqdef & a.C_1 + \epsilon.C_2 + \1\\
C_2 & \eqdef & b.C_2 + \epsilon.C_3\\
C_3 & \eqdef & a.C_3 + \1\\
\end{array}$

\noindent It is easily seen that $W_g \vdash C_3 = a.C_3 + \1$, 
$W_g \vdash C_2 = b.C_2 + a.C_3 + \1$ and so also $W_g \vdash C_1 = a.C_1 + b.C_2 + a.C_3 + \1$.
Therefore, we can define a new $\epsilon$-free system of equations:

$\begin{array}{rcl}
D_1 & \eqdef & a.D_1 + b.D_2 + a.D_3 + \1\\
D_2 & \eqdef & b.D_2 + a.D_3 + \1\\
D_3 & \eqdef & a.D_3 + \1\\
\end{array}$

\noindent
Finally, by Theorem \ref{unique-wg-th}, we get $W_g \vdash C_i = D_i$ for $i= 1, 2, 3$.
\fine
\end{example}

\begin{proposition}\label{eps-free-prop}{\bf (Reduction to $\epsilon$-free normal form)}
Let $\widetilde{X} = (x_1, x_2, \ldots, x_n)$ be a tuple of variables. Let 
$\widetilde{p} = (p_1, p_2, \ldots, p_n)$ be a tuple of
open guarded SFM1 terms, using the variables in $\widetilde{X}$, of the form $p_h = \sum_{i = 1}^{m(h)} \alpha_{hi}.x_{f(h,i)} \{+\1\}$
for $h = 1, \ldots, n$ and where $f$ is such that $1 \leq f(h,i) \leq n$.
Let $\widetilde{C} = (C_1, C_2, \ldots, C_n)$ be a tuple of constants 
such that the system of equations in normal form
 
 $\begin{array}{lcllcllcl}
C_1 & \eqdef &  p_1\{\widetilde{C}/\widetilde{X}\}\\
C_2 & \eqdef & p_2\{\widetilde{C}/\widetilde{X}\}\\
\ldots\\
C_n & \eqdef & p_n\{\widetilde{C}/\widetilde{X}\}
\end{array}$\\

\noindent
is observationally guarded. Then, there exist a tuple $\widetilde{q} = (q_1, q_2, \ldots, q_n)$ of
open guarded SFM1 terms, using the variables in $\widetilde{X}$, and a tuple 
of constants $\widetilde{D} = (D_1, D_2, \ldots, D_n)$  such that the system of equations in normal form
 
 $\begin{array}{lcllcllcl}
D_1 & \eqdef &  q_1\{\widetilde{D}/\widetilde{X}\}\\
D_2 & \eqdef & q_2\{\widetilde{D}/\widetilde{X}\}\\
\ldots\\
D_n & \eqdef & q_n\{\widetilde{C}/\widetilde{X}\}
\end{array}$\\

\noindent
is $\epsilon$-free and, for $i = 1, \ldots, n$, 
$W_g \vdash C_i = D_i$.

\proof By induction on $n$. For $n = 1$, we have that $C_1 \eqdef p_1\{C_1/x_1\}$,
where $p_1$ must be of the form
$\sum_{i = 1}^{m(1)} \alpha_{1i}.x_1 \{+\1\}$. Since $p_1$ is observationally guarded, all the $\alpha_{1i}$ must be in $A$,
so that $C_1$ is already $\epsilon$-free.

Now assume a tuple $\widetilde{p} = (p_1, p_2, \ldots, p_n)$ and the term $p_{n+1}$,
so that they are all open on $\widetilde{X} = (x_1, x_2, \ldots, x_n)$ and the additional $x_{n+1}$.
Assume, w.l.o.g., that $x_{n+1}$ occurs in $p_{n+1}$. 
First, define 
$C_{n+1} \eqdef p_{n+1}\{C_{n+1}/x_{n+1}\}$, so that $C_{n+1}$ is now open on $\widetilde{X}$.
Therefore, also for $i = 1, \ldots, n$, each $p_{i}\{C_{n+1}/x_{n+1}\}$ is now
open on $\widetilde{X}$.
The resulting observationally guarded
system of equations in normal form of size $n$ is:

$\begin{array}{lcllcllcl}
C_1 & \eqdef &  p_1\{C_{n+1}/x_{n+1}\}\{\widetilde{C}/\widetilde{X}\}\\
C_2 & \eqdef & p_2\{C_{n+1}/x_{n+1}\}\{\widetilde{C}/\widetilde{X}\}\\
\ldots\\
C_n & \eqdef & p_n\{C_{n+1}/x_{n+1}\}\{\widetilde{C}/\widetilde{X}\}\\
\end{array}$

\noindent 
so that, by induction, we can conclude that
there exist a tuple $\widetilde{q} = (q_1, q_2, \ldots, q_n)$ of terms and a tuple of constants $\widetilde{D} = (D_1, D_2, \ldots, D_n)$
such that 

$\begin{array}{lcllcllcl}
D_1 & \eqdef &  q_1\{C_{n+1}/x_{n+1}\}\{\widetilde{D}/\widetilde{X}\}\\
D_2 & \eqdef & q_2\{C_{n+1}/x_{n+1}\}\{\widetilde{D}/\widetilde{X}\}\\
\ldots\\
D_n & \eqdef & q_n\{C_{n+1}/x_{n+1}\}\{\widetilde{D}/\widetilde{X}\}
\end{array}$\\

\noindent 
is $\epsilon$-free and $W_g \vdash C_i = D_i$ for $i = 1, \ldots, n$.
Note that 

$q_i\{C_{n+1}/x_{n+1}\}\{\widetilde{D}/\widetilde{X}\} = q_{i}\{\widetilde{D}/\widetilde{X}, C_{n+1}\{\widetilde{D}/\widetilde{X}\}/x_{n+1}\}$

\noindent
so that by $C_{n+1}\{\widetilde{D}/\widetilde{X}\}$ we have implicitly closed the definition of 
$C_{n+1}$ as
 
$C_{n+1} \eqdef p_{n+1}\{C_{n+1}/x_{n+1}\}\{\widetilde{D}/\widetilde{X}\} = 
p_{n+1}\{\widetilde{D}/\widetilde{X}\}\{C_{n+1}/x_{n+1}\}$.

Note that inside the term $p_{n+1}\{\widetilde{D}/\widetilde{X}\}$ there cannot be
any summand of the form $\epsilon.x_{n+1}$ because the system is observationally guarded.
Now we unfold each constant $D_i$ inside the term $p_{n+1}\{\widetilde{D}/\widetilde{X}\}$
which is prefixed by $\epsilon$, so that each summand of the form $\epsilon.D_i$ is replaced by the summand $body(D_i)$,
resulting in the $\epsilon$-free term $q_{n+1}\{\widetilde{D}/\widetilde{X}\}$
Note that, by axioms $\{${\bf T3,R1}$\}$, we have

$W_g \vdash p_{n+1}\{\widetilde{D}/\widetilde{X}\} =  q_{n+1}\{\widetilde{D}/\widetilde{X}\}$.

Now define a new constant $D_{n+1} \eqdef q_{n+1}\{\widetilde{D}/\widetilde{X}\}\{D_{n+1}/x_{n+1}\}$ so that
$W_g \vdash C_{n+1} = D_{n+1}$ by recursion congruence.
Moreover, by axiom {\bf R1}, we have that 
$W_g \vdash D_{n+1} = q_{n+1}\{D_{n+1}/x_{n+1}\}\{\widetilde{D}/\widetilde{X}\}$. 
Similarly, also that
$W_g \vdash D_i  =  q_i\{C_{n+1}/x_{n+1}\}\{\widetilde{D}/\widetilde{X}\}$ for $i = 1, \ldots, n$. By substitutivity, we also get
$W_g \vdash D_i  =  q_i\{D_{n+1}/x_{n+1}\}\{\widetilde{D}/\widetilde{X}\}$ for $i = 1, \ldots, n$.
Finally, 
we define a new $\epsilon$-free system of equations in normal form

$\begin{array}{lcllcllcl}
E_1 & \eqdef &  q_1\{E_{n+1}/x_{n+1}\}\{\widetilde{E}/\widetilde{X}\}\\
E_2 & \eqdef & q_2\{E_{n+1}/x_{n+1}\}\{\widetilde{E}/\widetilde{X}\}\\
\ldots\\
E_n & \eqdef & q_n\{E_{n+1}/x_{n+1}\}\{\widetilde{E}/\widetilde{X}\}\\
E_{n+1} & \eqdef & q_{n+1}\{E_{n+1}/x_{n+1}\}\{\widetilde{E}/\widetilde{X}\}
\end{array}$\\

\noindent
such that $W_g \vdash D_i  = E_i$ for $i = 1, \ldots, n+1$ by Theorem \ref{unique-wg-th}. The thesis, $W_g \vdash C_i  = E_i$ for $i = 1, \ldots, n+1$, 
follows by transitivity.
\fine
\end{proposition}

\subsection{Deterministic Normal Form}\label{det-nform-sec}

Given an alphabet $A = \{a_1, \ldots, a_k\}$, we say that an $\epsilon$-free normal form is deterministic if it is of the form\\

$\begin{array}{lcllcllcl}
C_1 & \eqdef &  \sum_{i = 1}^{k} a_{i}.C_{f(1,i)} \{+\1\}\\
C_2 & \eqdef &  \sum_{i = 1}^{k} a_{i}.C_{f(2,i)} \{+\1\}\\
\ldots\\
C_n & \eqdef & \sum_{i = 1}^{k} a_{i}.C_{f(n,i)} \{+\1\}\\
\end{array}$\\

\noindent
i.e., the body of each constant contains exactly one summand for each action $a_i \in A$. Of course,
the NFA associated to $C_1$, according to the denotational semantics in Table \ref{den-nfa-sfm1}, is deterministic 
and has exactly $A$ as its alphabet.

Note that the concept of determinism heavily depends on the chosen alphabet. 
For instance, according to the denotational semantics in Table \ref{den-nfa-sfm1}, the NFA for $C \eqdef \nil$ is deterministic, because its alphabet is empty. 
However, it is not deterministic if we consider 
a nonempty alphabet, say $A = \{a\}$,
because there is not an arc for $a$. In such a case, a DFA w.r.t. $A$, language equivalent to $C \eqdef \nil$, could be the one associated to
$E \eqdef a.E$. Indeed, it is possible to prove that $W_g \vdash C = E$ by means of axioms {\bf T1, R1, R2}; in fact, 
by axiom {\bf T1}, we have $W_g \vdash \nil = a.\nil$; then, by axiom {\bf R2}, from $E \eqdef a.E$ and 
$W_g \vdash \nil = a.\nil$, we get $W_g \vdash E = \nil$; then by axiom {\bf R1}, we have $W_g \vdash C = \nil$, so that
$W_g \vdash C = E$ follows by transitivity.

Similarly, the NFA for $C \eqdef \1$ is deterministic, because its alphabet is empty. 
However, it is not deterministic if we consider 
the nonempty alphabet $A = \{a\}$. In such a case, a DFA w.r.t. $A$, language equivalent to $C \eqdef \1$, could be the one associated to
$D \eqdef a.E + \1$, where $E \eqdef a.E$ denotes a sink, i.e., an error state. We can prove that $W_g \vdash C = D$ as follows.
First, we already know that $W_g \vdash E = \nil$; so, by substitutivity and {\bf R1}, we have $W_g \vdash D = a.\nil + \1$; then by {\bf T1},
$W_g \vdash D = \nil + \1$; then by axioms {\bf A2,A3}, we have $W_g \vdash D = \1$, so that $W_g \vdash C = D$ follows easily.

Now we prove that, given an $\epsilon$-free normal form, it is possible to construct a provably equivalent deterministic normal form.
The idea behind the proof, that follows the classic Rabin-Scott {\em subset construction} \cite{RS59}, is illustrated by the following example.

\begin{example}\label{ex-det}
Let us consider the following $\epsilon$-free normal form

$\begin{array}{rcl}
C_1 & \eqdef & a.C_1 +  a.C_2 +  a.C_1\\
C_2 & \eqdef & a.C_2 + \1\\
\end{array}$\\

\noindent
whose alphabet $A_C$ is $\{a\}$. Let us take an alphabet $A = \{a, b\}$, w.r.t. which we want to define an equivalent 
deterministic normal form. First, we define a collection of set-indexed 
constants $B_I$, where $I \subseteq \{1, 2\}$, as $B_I \eqdef \sum_{i \in I} body(C_i)$. More explicitly,

$\begin{array}{rcl}
B_{\{1\}} & \eqdef & a.C_1 +  a.C_2 +  a.C_1\\
B_{\{2\}} & \eqdef & a.C_2 + \1\\
B_{\{1,2\}} & \eqdef & a.C_1 +  a.C_2 +  a.C_1 + a.C_2 + \1\\
B_\emptyset & \eqdef & \nil\\
\end{array}$\\

\noindent
Now, for each of them, we prove that it can be equated to a deterministic definition. First, we can prove that, 
since $B_\emptyset = \nil$ by axiom {\bf R1}, also
$W_g \vdash B_\emptyset = a.B_{\emptyset} + b.B_{\emptyset}$ by axioms {\bf T1,A3}.

Now, by axiom {\bf R1}, we know that $W_g \vdash B_{\{1\}} = a.C_1 +  a.C_2 +  a.C_1$, so 
that $W_g \vdash C_1 = B_{\{1\}}$. Then, by axioms {\bf A1-A4,T2,R1},
we can derive $W_g \vdash B_{\{1\}} = a.(body(C_1) + body(C_2))$. Then, we can also derive 
$W_g \vdash B_{\{1\}} = a.B_{\{1,2\}} + b.B_\emptyset$ by axioms {\bf R1,T1,A3}. 

In the same way, we can prove that $W_g \vdash B_{\{2\}} = a.B_{\{2\}} + b.B_\emptyset + \1$ and that
$W_g \vdash B_{\{1,2\}} = a.B_{\{1,2\}} + b.B_\emptyset + \1$.

Finally, we define a new deterministic system of equations

$\begin{array}{rcl}
D_{\{1\}} & \eqdef & a.D_{\{1,2\}} + b.D_{\emptyset}\\
D_{\{2\}} & \eqdef & a.D_{\{2\}} + b.D_{\emptyset} + \1\\
D_{\{1,2\}} & \eqdef &  a.D_{\{1,2\}} + b.D_{\emptyset} + \1\\
D_{\emptyset} & \eqdef & a.D_{\emptyset} + b.D_{\emptyset}\\
\end{array}$\\

\noindent
where $D_{\{2\}}$ is redundant as $D_{\{2\}} \not \in \const{D_{\{1\}}}$, such that, by Theorem \ref{unique-wg-th}, we have $W_g \vdash B_I = D_I$ for $I = \{1\}, \{2\}, \{1,2\}, \emptyset$.
So, $W_g \vdash C_1 = D_{\{1\}}$ by transitivity.
\fine
\end{example}

\begin{proposition}\label{det-red-prop}{\bf (Reduction to deterministic normal form)}
Given the $\epsilon$-free normal form
 
$\begin{array}{lcllcllcl}
C_1 & \eqdef & \sum_{i = 1}^{m(1)} a_{1i}.C_{f(1,i)} \{+\1\} \\
C_2 & \eqdef &  \sum_{i = 1}^{m(2)} a_{2i}.C_{f(2,i)} \{+\1\} \\
\ldots\\
C_n & \eqdef & \sum_{i = 1}^{m(n)} a_{ni}.C_{f(n,i)} \{+\1\} \\
\end{array}$\\

\noindent 
such that its alphabet is $A_{C}$,
it is possible to construct a deterministic normal form w.r.t. a chosen $A = \{a_1, a_2, \ldots, a_k\}$,
with $A_C \subseteq A$, of the form 

$\begin{array}{lcllcllcl}
D_1 & \eqdef &  \sum_{i = 1}^{k} a_{i}.D_{g(1,i)} \{+\1\}\\
D_2 & \eqdef &  \sum_{i = 1}^{k} a_{i}.D_{g(2,i)} \{+\1\}\\
\ldots\\
D_m & \eqdef & \sum_{i = 1}^{k} a_{i}.D_{g(m,i)} \{+\1\}\\
\end{array}$\\

\noindent
where $m \geq n$, such that $W_g \vdash C_1 = D_1$.

\proof 
We define
a set of set-indexed constants $B_I \eqdef \sum_{i \in I} body(C_i)$, one for each $I \subseteq \{1, \ldots, n\}$.
Note that $W_g \vdash C_1 = B_{\{1\}}$ by axiom {\bf R1}.
Now for each of these constants $B_I$ we want to show that it can be equated to a deterministic definition, i.e., a definition 
where there is exactly one summand per symbol in $A$. First, let $In(B_I)$ denote the set of initial symbols of $B_I$, i.e.,
$\cup_{i \in I} \{a_{i\,1}, \ldots, a_{i\, m(i)}\}$. Then, 
by using the axioms {\bf A1-A4,R1,T2}  and substitutivity, we can prove that

$W_g \vdash B_I = \sum_{a_{h}\in In(B_I)} a_{h}.(\sum_{i \in I_h} body(C_i))$,

\noindent
where $I_h = \{f(i,j) \mid a_{i,j} = a_h, a_{i,j} \in In(B_I)\}$.
The next step is to recognize that $\sum_{i \in I_h} body(C_i)$ can be replaced by $B_{I_h}$ by axiom {\bf R1}, so that

$W_g \vdash B_I = \sum_{a_{h}\in In(B_I)} a_{h}.B_{I_h}$.

\noindent
The next step is to add summands related to those symbols in $A$ not in $In(B_I)$. First note that $B_\emptyset \eqdef \nil$ 
can be equated, by means of axioms {\bf R1,T1,A3}, to 

$W_g \vdash B_\emptyset = \sum_{a_i \in A} a_i.B_\emptyset$, 

\noindent 
which is a deterministic definition for the error state $B_\emptyset$. Therefore, by axioms {\bf T1,A3} we have

$W_g \vdash B_I = \sum_{a_{h}\in In(B_I)} a_{h}.B_{I_h} + \sum_{a_h \not\in In(B_I)} a_h.B_\emptyset$

\noindent 
which is a deterministic definition for constant $B_I$. Finally, we can define a deterministic system of equations 
of the form

$D_I \eqdef \sum_{a_{h}\in In(B_I)} a_{h}.D_{I_h} + \sum_{a_h \not\in In(B_I)} a_h.D_\emptyset$

\noindent
for each $I \subseteq \{1, \ldots, n\}$, such that, by Theorem \ref{unique-wg-th}, we have $W_g \vdash B_I = D_I$. 
So, $W_g \vdash C_1 = D_{\{1\}}$ by transitivity.
\fine
\end{proposition}

\subsection{Completeness for Observationally Guarded Processes}\label{compl-obg-sec}

Proposition \ref{det-red-prop} is at the basis of the lemma below, which, in turn, is crucial for the proof of completeness
for observationally guarded processes that follows. The proof technique of the lemma below is inspired to the proof
of Theorem 5.10 in \cite{Mil84}, in turn inspired to the proof of Theorem 2 in \cite{Salomaa}.

\begin{lemma}{\bf (Completeness for $\epsilon$-free normal forms)}\label{det-nf-compl}
For every $p, p' \in  \mathcal{P}_{SFM1}$ $\epsilon$-free normal forms, 
if $p \sim p'$, then $W_g \vdash p = p'$.

\proof Let $A_p$ the alphabet of $p$ and $A_{p'}$ the alphabet of $p'$. Take $A = A_{p} \cup A_{p'}$.
By Proposition \ref{det-red-prop}, there exist $p_1$ and $p_1'$ deterministic normal forms w.r.t. $A$,
such that $W_g \vdash p = p_1$ and $W_g \vdash p' = p_1'$. By Theorem \ref{sound-th-sfm1}, we have $p \sim p_1$
and $p' \sim p_1'$ and so, by transitivity, also $p_1 \sim p_1'$.

Let $A = \{a_1, \ldots, a_k\}$. We can assume that $p_1$
is the deterministic system of equations: 

$\begin{array}{lcllcllcl}
C_1 & \eqdef &  \sum_{i = 1}^{k} a_{i}.C_{f(1,i)} \{+\1\}\\
C_2 & \eqdef &  \sum_{i = 1}^{k} a_{i}.C_{f(2,i)} \{+\1\}\\
\ldots\\
C_n & \eqdef & \sum_{i = 1}^{k} a_{i}.C_{f(n,i)} \{+\1\}\\
\end{array}$\\

\noindent 
so that $p_1 = C_1$. For each $h = 1, \ldots, n$,
we get
$W_g \vdash  C_h = body(C_h)$, by axiom {\bf R1}, where by $body(C_h)$ we denote 
$\sum_{i = 1}^{k} a_{i}.C_{f(h,i)} \{+\1\}$.

Similarly, we can assume that $p_1'$ is the deterministic system of 
equations: 

$\begin{array}{lcllcllcl}
C'_1 & \eqdef &  \sum_{i = 1}^{k} a_{i}.C'_{f'(1,i)} \{+\1\}\\
C'_2 & \eqdef &  \sum_{i = 1}^{k} a_{i}.C'_{f'(2,i)} \{+\1\}\\
\ldots\\
C'_{n'} & \eqdef & \sum_{i = 1}^{k} a_{i}.C'_{f'(n',i)} \{+\1\}\\
\end{array}$\\

\noindent 
so that $p_1' = C'_1$. For each $h' = 1, \ldots, n'$, 
we get
$W_g \vdash  C'_{h'} = body(C'_{h'})$, by axiom {\bf R1}, where by $body(C'_{h'})$ we denote 
$\sum_{i = 1}^{k} a_{i}.C'_{f'(h',i)} \{+\1\}$.
Moreover, as $p_1 \sim p_1'$, we have $C_1 \sim C'_1$. 

Now, let $H = \{(h, h') \mid $  
$ C_h \sim C'_{h'}\}$. Of course, note that $(1,1) \in H$.
Moreover, for $(h, h') \in H$, since  $C_h$ and $C'_{h'}$ are language equivalent and deterministic (hence, bisimulation equivalent, 
cf. Remark \ref{lang=bis-rem}), the following hold:
\begin{itemize}
\item for each $i = 1, \ldots, k$, we have $(f(h,i), f'(h',i)) \in H$, and
\item the optional summand $\1$ belongs to $body(C_h)$ iff it belongs to $body(C'_{h'})$.
\end{itemize}

\noindent
Now, for each $(h, h') \in H$, let us consider the open terms 

$
t_{hh'} = \sum_{i = 1}^{k}  a_{i}.x_{f(h,i),f'(h',i)} \{+\1\}
$

\noindent
By Theorem \ref{unique-wg-th}, for each $(h, h') \in H$,  there exists a constant $D_{hh'}$ such that $W_g \vdash D_{hh'} 
= t_{hh'}\{\widetilde{D}/\widetilde{X}\} $, where $\widetilde{D}$ denotes the tuple of constants $D_{hh'}$ 
for each $(h, h') \in H$, and $\widetilde{X}$ denotes the tuple of variables $x_{hh'}$ for each $(h, h') \in H$.
More explicitly,
$W_g \vdash  D_{hh'} = \sum_{i = 1}^{k}  a_{i}.D_{f(h,i),f'(h',i)}  \{+\1\}$.
If we close each $t_{hh'}$ by replacing $x_{f(h,i),f'(h',i)}$ with $C_{f(h,i)}$, we get

$
\quad \sum_{i = 1}^{k}  a_{i}.C_{f(h,i)}\{+\1\}
$

\noindent
which is equal to $body(C_h)$.
Therefore, we note that $C_h$  is 
such that $W_g \vdash C_h = t_{hh'}\{\widetilde{C}/\widetilde{X}\}$ and
so, by Theorem \ref{unique-wg-th}, we have that $W_g \vdash D_{hh'} = C_h$.
Since $(1, 1) \in H$, we have that $W_g \vdash D_{11} = C_1$.
Similarly, if we close each $t_{hh'}$ by replacing $x_{f(h,i),f'(h',i)}$ with $C'_{f'(h',i)}$, we get

$
\quad \sum_{i = 1}^{k}  a_{i}.C'_{f'(h',i)} \{+\1\}
$

\noindent
which is equal to $body(C'_{h'})$.  Therefore, we note that $C'_{h'}$ is 
such that $W_g \vdash C'_{h'} = t_{hh'}\{\widetilde{C'}/\widetilde{X}\}$ and
so, by Theorem \ref{unique-wg-th}, we have 
that $W_g \vdash D_{hh'} = C'_{h'}$. Since $(1, 1) \in H$, we have that  
$W_g \vdash D_{11} = C'_1$; by transitivity, it follows that $W_g \vdash C_1 = C'_1$, and so that $W_g \vdash p_1 = p_1'$.
Finally, the thesis $W_g \vdash p = p'$ follows by transitivity.
\fine
\end{lemma}

\begin{theorem}{\bf (Completeness for observationally guarded processes)}\label{compl-og-ax}

\noindent
For every $p, q \in  \mathcal{P}_{SFM1}$ observationally guarded, if $p \sim q$, then $W_g \vdash p = q$.

\proof
By Corollary \ref{nf-cor}, there exist $p_1$ and $q_1$ observationally guarded
normal forms such that $W_g \vdash p = p_1$ and $W_g \vdash q = q_1$. 
By Theorem \ref{sound-th-sfm1}, we have $p \sim p_1$ and $q \sim q_1$,
hence, by transitivity, also $p_1 \sim q_1$.
By Proposition \ref{eps-free-prop}, there exist $p_2$ and $q_2$ $\epsilon$-free normal forms, such that $W_g \vdash p_1 = p_2$
and $W_g \vdash q_1 = q_2$. By Theorem \ref{sound-th-sfm1}, we have $p_1 \sim p_2$ and $q_1 \sim q_2$, and so, by transitivity,
also $p_2 \sim q_2$.
By Lemma \ref{det-nf-compl}, from $p_2 \sim q_2$ we can derive that $W_g \vdash p_2 = q_2$, and so, by transitivity, also
$W_g \vdash p = q$.
\fine
\end{theorem}

\subsection{Completeness for Unguarded Processes}\label{det-nf-sec}

In the following we restrict our attention to  {\em systems of equations}, not necessarily in normal form.
In fact, if $p$ is a process and 
$p$ is not a constant, then we can take a new constant $D \eqdef p$ (such that $W \vdash D = p$
by axiom {\bf R1}), so that $p$ is equivalently represented by the system of equations over the set of constants \const{D}.

We now prove that the addition of axiom {\bf R3} to $W_g$ is enough to 
equate any SFM1 process $p$
to an observationally guarded SFM1 process $q$. 
In order to prove this result, we need some auxiliary notation.

First of all, we denote by $n_x(p)$ the number of unguarded occurrences of $x$ in $p$ (including
the bodies of the constants occurring in $p$). E.g., if $C \eqdef p\{C/x\}$ and 
$p = \epsilon.x + a.x + \epsilon.D$, with $D \eqdef \epsilon.D + \epsilon.x$, then $n_x(p) = 2$; moreover, we set
$n_x(C) = n_x(p)$.
 
Then, we need to introduce a measure of the length of all the $\epsilon$-labeled computations leading to 
an occurrence of an unguarded
variable $x$. This value $\len(p)$ for the SFM1 term $p$ (which may be open on $x$,
but the definition applies also to closed terms)
is computed by the length function defined in Table \ref{len-tab}.
Note that if a symbol, say $a$, of the alphabet prefixes a process, then the returned value is $0$,
while the variable $x$ returns $1$. Moreover, in case of composition with the choice operator,
$p + p'$, we take the sum of $\len(p)$ and $\len(p')$.
The crucial rule is that for the constant: if $C \eqdef p\{C/x\}$, then, even if $C$ is closed,
its length is computed on the open term $p$. 
It is an easy
exercise to show that the open term $p$ is observationally guarded (cf. Definition \ref{og-def}) if 
and only if $\len(p) = n_x(p) = 0$. Moreover, if $p$ is closed (actually a closed system of equations), 
then $p$ is observationally guarded (cf. Remark \ref{rem-obs-g}) if and only if $\len(p) = n_x(p) = 0$.

\begin{table}[!t]

{\renewcommand{\arraystretch}{1.2}
\hrulefill\\[-.7cm]
			\begin{center}\fontsize{8pt}{0.1pt}
$\begin{array}{rcllrcllllll}
\len(\1) & = & 0 & \; & \len(\nil) & = & 0 \\
 \len(p + p') & = &  \len(p) + \len(p') & \; & \len(C) & = & 
  \len(p)  \quad \mbox{if $C \eqdef p\{C/x\}$} \\
\len(x) & = & 1 & \; &
\len(\epsilon.p) & = & \begin{cases}
  1 + \len(p) & \mbox{if $\len(p) \neq 0$}\\  
  0  & \mbox{otherwise}   
   \end{cases}\\
\len(a.p) & = &  0 & \; &\\[-.2cm]
\end{array}$
\end{center}}
\hrulefill
\caption{Length function} \protect\label{len-tab}
\end{table}

\begin{proposition}\label{ung-sfm1-prop}{\bf (Reduction to observationally guarded process for $W$)}
Let $\widetilde{X} = (x_1, x_2, \ldots, x_n)$ be a tuple of variables and let 
$\widetilde{p} = (p_1, p_2, \ldots, p_n)$ be a tuple of
open guarded SFM1 terms, using the variables in $\widetilde{X}$. 
Let $\widetilde{C} = (C_1, C_2, \ldots, C_n)$ be a tuple of constants (not occurring in $\widetilde{p}$) 
such that the system of equations
 
 $\begin{array}{lcllcllcl}
C_1 & \eqdef &  p_1\{\widetilde{C}/\widetilde{X}\}\\
C_2 & \eqdef & p_2\{\widetilde{C}/\widetilde{X}\}\\
\ldots\\
C_n & \eqdef & p_n\{\widetilde{C}/\widetilde{X}\}
\end{array}$\\

\noindent
is observationally unguarded. Then, there exist a tuple $\widetilde{q} = (q_1, q_2, \ldots, q_n)$ of
open guarded SFM1 terms, using the variables in $\widetilde{X}$, and a tuple 
 $\widetilde{D} = (D_1, D_2, \ldots, D_n)$  (not occurring in $\widetilde{q}$)  such that the system of equations
 
 $\begin{array}{lcllcllcl}
D_1 & \eqdef &  q_1\{\widetilde{D}/\widetilde{X}\}\\
D_2 & \eqdef & q_2\{\widetilde{D}/\widetilde{X}\}\\
\ldots\\
D_n & \eqdef & q_n\{\widetilde{C}/\widetilde{X}\}
\end{array}$\\

\noindent
is observationally guarded and, for $i = 1, \ldots, n$, 
$W \vdash C_i = D_i$.

\proof 
The proof is by double induction: first on $n$, and then on the pair $(n_x(p), \len(p))$ (for the considered open guarded term $p$), 
where we assume that $(n_1, k_2) < (n_2, k_2)$
if $n_1 < n_2$, or $n_1 = n_2$ and $k_1 < k_2$.

For $n = 1$, if $(n_{x_1}(p_1), \len(p_1)) = (0, 0)$, then we are done, as $C_1 \eqdef p_1\{C_1/x_{1}\}$ is actually observationally guarded.
Instead, if $p_1$ is observationally unguarded, then it must be of the form $\epsilon.r + q$, with
$r$ observationally unguarded. We now proceed by case analysis: 
\begin{itemize}
\item 
$r = x_1$: In this case, $WB \vdash C_1 = (\epsilon.x_1 + q)\{C_1/x_1\}$ by axiom {\bf R1} (and possibly also {\bf A1-A4}).
Now define $C_2 \eqdef (\epsilon.x_1 + q)\{C_2/x_1\}$ so that $W \vdash C_1 = C_2$ by recursion congruence.
Define $C_3 \eqdef q\{C_3/x_1\}$: by axiom {\bf R3} we have $W \vdash C_2 = C_3$. 
Note that $n_{x_1}(q) = n_{x_1}(\epsilon.x_1 + q)-1$, so that induction 
can be invoked to
conclude that there exist an observationally guarded process $q_1$ and a constant $D_1$ such that 
$D_1 \eqdef q_1\{D_1/x_1\}$ and $W \vdash 
C_3 = D_1$, so that $W \vdash C_1 = D_1$ follows by transitivity.

\item
$r = \epsilon.p_1 + p_2$, with $p_1$ observationally unguarded. 
Thus, $W \vdash C_1 =  (\epsilon.(\epsilon.p_1 + p_2) + q)\{C_1/x_1\}$ by axiom {\bf R1} (and possibly also {\bf A1-A4}).
Since by axiom {\bf T3} we have $W \vdash \epsilon.(\epsilon.p_1 + p_2) = \epsilon.p_1 + p_2$, by substitutivity we also have
$W \vdash C_1 =  ((\epsilon.p_1 + p_2) + q)\{C_1/x_1\}$. 
Now define $C_2 \eqdef ((\epsilon.p_1 + p_2) + q)\{C_2/x_1\}$ so that $W \vdash C_1 = C_2$ by recursion congruence.
Note that $\len((\epsilon.p_1 + p_2) + q) < \len(\epsilon.(\epsilon.p_1 + p_2) + q)$,
so that induction can be invoked to
conclude that there exist an observationally guarded process $q_1$ and a constant $D_1$ such that
$D_1 \eqdef q_1\{D_1/x\}$ and $W \vdash C_2 = D_1$, so that $W \vdash C_1 = D_1$ follows by transitivity.
\end{itemize}
As no other cases are possible, the proof of the base case for $n = 1$ is complete.
 
Now assume a tuple $\widetilde{p} = (p_1, p_2, \ldots, p_n)$ and the term $p_{n+1}$,
so that they are all open on $\widetilde{X} = (x_1, x_2, \ldots, x_n)$ and the additional $x_{n+1}$.
Assume, w.l.o.g., that $x_{n+1}$ occurs in $p_{n+1}$. 
First, define 

$C_{n+1} \eqdef p_{n+1}\{C_{n+1}/x_{n+1}\}$, 

\noindent
so that $C_{n+1}$ is now open on $\widetilde{X}$.
Then, we define $p_{i}\{C_{n+1}/x_{n+1}\}$, which, for $i = 1, \ldots, n$, is the term
$p_i$ where each occurrence of variable $x_{n+1}$ has been replaced by  $C_{n+1}$;
note that each term $p_{i}\{C_{n+1}/x_{n+1}\}$ is open on $\widetilde{X}$.
Hence\\

 $\begin{array}{lcllcllcl}
C_1 & \eqdef &  p_1\{C_{n+1}/x_{n+1}\}\{\widetilde{C}/\widetilde{X}\}\\
C_2 & \eqdef & p_2\{C_{n+1}/x_{n+1}\}\{\widetilde{C}/\widetilde{X}\}\\
\ldots\\
C_n & \eqdef & p_n\{C_{n+1}/x_{n+1}\}\{\widetilde{C}/\widetilde{X}\}
\end{array}$\\

\noindent
is a system of equation of size $n$, so that, by induction, we can conclude that
there exist a tuple $\widetilde{q} = (q_1, q_2, \ldots, q_n)$ of terms and a tuple of constants $\widetilde{D} = (D_1, D_2, \ldots, D_n)$
such that 

$\begin{array}{lcllcllcl}
D_1 & \eqdef &  q_1\{C_{n+1}/x_{n+1}\}\{\widetilde{D}/\widetilde{X}\}\\
D_2 & \eqdef & q_2\{C_{n+1}/x_{n+1}\}\{\widetilde{D}/\widetilde{X}\}\\
\ldots\\
D_n & \eqdef & q_n\{C_{n+1}/x_{n+1}\}\{\widetilde{D}/\widetilde{X}\}
\end{array}$\\

\noindent 
is observationally guarded and $W \vdash C_i = D_i$ for $i = 1, \ldots, n$.
Note that 

$q_i\{C_{n+1}/x_{n+1}\}\{\widetilde{D}/\widetilde{X}\} = q_{i}\{\widetilde{D}/\widetilde{X}, C_{n+1}\{\widetilde{D}/\widetilde{X}\}/x_{n+1}\}$

\noindent
so that by $C_{n+1}\{\widetilde{D}/\widetilde{X}\}$ we have implicitly closed the definition of 
$C_{n+1}$ as
 
$C_{n+1} \eqdef p_{n+1}\{C_{n+1}/x_{n+1}\}\{\widetilde{D}/\widetilde{X}\} = 
p_{n+1}\{\widetilde{D}/\widetilde{X}\}\{C_{n+1}/x_{n+1}\}$.

Now we unfold each constant $D_i$ inside the term $p_{n+1}\{\widetilde{D}/\widetilde{X}\}\{C_{n+1}/x_{n+1}\}$
so that possible further occurrences of $C_{n+1}$ are now exposed.
Note that, by axiom {\bf R1}, 

$WB \vdash p_{n+1}\{\widetilde{D}/\widetilde{X}\} = p_{n+1}\{\widetilde{body(D)}/\widetilde{X}\}$.

More compactly, we can write 

$p_{n+1}\{\widetilde{body(D)}/\widetilde{X}\} = r_{n+1}\{\widetilde{D}/\widetilde{X}\}$.

\noindent
and define a new constant $C'_{n+1} \eqdef r_{n+1}\{\widetilde{D}/\widetilde{X}\}\{C'_{n+1}/x_{n+1}\}$ so that
$W \vdash C_{n+1} = C'_{n+1}$ by recursion congruence.
Note that we are now in a situation similar to the base case for $n=1$: 
\begin{itemize}
\item
if $(n_{x_{n+1}}(r_{n+1}\{\widetilde{D}/\widetilde{X}\}),$ $\len(r_{n+1}\{\widetilde{D}/\widetilde{X}\})) = (0, 0)$, then we are (almost) 
done, as 
$C'_{n+1} \eqdef r_{n+1}\{\widetilde{D}/\widetilde{X}\}\{C'_{n+1}/x_{n+1}\}$ is actually observationally guarded.
By axiom {\bf R1}, we have that 
$W \vdash C'_{n+1} = r_{n+1}\{C'_{n+1}/x_{n+1}\}\{\widetilde{D}/\widetilde{X}\}$. 

Similarly, also that
$W \vdash D_i  =  q_i\{C_{n+1}/x_{n+1}\}\{\widetilde{D}/\widetilde{X}\}$ for $i = 1, \ldots, n$. By substitutivity, we also get
$W \vdash D_i  =  q_i\{C'_{n+1}/x_{n+1}\}\{\widetilde{D}/\widetilde{X}\}$ for $i = 1, \ldots, n$.
Finally, 
we define a new observationally guarded system of equations

$\begin{array}{lcllcllcl}
E_1 & \eqdef &  q_1\{E_{n+1}/x_{n+1}\}\{\widetilde{E}/\widetilde{X}\}\\
E_2 & \eqdef & q_2\{E_{n+1}/x_{n+1}\}\{\widetilde{E}/\widetilde{X}\}\\
\ldots\\
E_n & \eqdef & q_n\{E_{n+1}/x_{n+1}\}\{\widetilde{E}/\widetilde{X}\}\\
E_{n+1} & \eqdef & r_{n+1}\{E_{n+1}/x_{n+1}\}\{\widetilde{E}/\widetilde{X}\}
\end{array}$\\

\noindent
such that $W \vdash D_i  = E_i$ for $i = 1, \ldots, n$ and $W \vdash C'_{n+1} = E_{n+1}$ by Theorem \ref{unique-wg-th}. The thesis, $W \vdash C_i  = E_i$ for $i = 1, \ldots, n+1$, 
follows by transitivity.

\item
Instead, if $r_{n+1}\{\widetilde{D}/\widetilde{X}\}$ is observationally unguarded, then it must be of the form $\epsilon.r + q$, with
$r$ observationally unguarded. We can now proceed by case analysis as done for the base case; as this is indeed very similar, we omit
this part of the proof. So, at the end, we get a new observationally guarded term $q_{n+1}$ and a new constant $D_{n+1}$ such that
$D_{n+1} \eqdef q_{n+1}\{D_{n+1}/x_{n+1}\}\{\widetilde{D}/\widetilde{X}\}$ and $W \vdash C'_{n+1} = D_{n+1}$. By transitivity,
$W \vdash C_{n+1} = D_{n+1}$. Moreover,
by axiom {\bf R1}, we have that 
$W \vdash D_{n+1} = q_{n+1}\{D_{n+1}/x_{n+1}\}\{\widetilde{D}/\widetilde{X}\}$. 

Similarly, also that
$W \vdash D_i  =  q_i\{C_{n+1}/x_{n+1}\}\{\widetilde{D}/\widetilde{X}\}$ for $i = 1, \ldots, n$. By substitutivity, we also get
$W \vdash D_i  =  q_i\{D_{n+1}/x_{n+1}\}\{\widetilde{D}/\widetilde{X}\}$ for $i = 1, \ldots, n$.
Finally, 
we define a new observationally guarded system of equations

$\begin{array}{lcllcllcl}
E_1 & \eqdef &  q_1\{E_{n+1}/x_{n+1}\}\{\widetilde{E}/\widetilde{X}\}\\
E_2 & \eqdef & q_2\{E_{n+1}/x_{n+1}\}\{\widetilde{E}/\widetilde{X}\}\\
\ldots\\
E_n & \eqdef & q_n\{E_{n+1}/x_{n+1}\}\{\widetilde{E}/\widetilde{X}\}\\
E_{n+1} & \eqdef & q_{n+1}\{E_{n+1}/x_{n+1}\}\{\widetilde{E}/\widetilde{X}\}
\end{array}$\\

\noindent
such that $W \vdash D_i  = E_i$ for $i = 1, \ldots, n+1$ by Theorem \ref{unique-wg-th}. The thesis, $W \vdash C_i  = E_i$ for $i = 1, \ldots, n+1$, 
follows by transitivity.\\[-1cm]
\end{itemize}
\fine
\end{proposition}

\begin{theorem}\label{compl-ung-w}{\bf (Completeness of $W$ over the whole of SFM1)}
For every $p, q \in  \mathcal{P}_{SFM1}$, if $p \sim q$, 
then $W \vdash p = q$.

\proof By Proposition \ref{ung-sfm1-prop}, there exists $p'$ and $q'$, observationally guarded, such that
$W \vdash p = p'$ and $W \vdash q = q'$. By Theorem \ref{sound-th-sfm1}, we have that
$p \sim p'$ as well as $q \sim q'$, so that, by transitivity, we also have
$p' \sim q'$. By Theorem \ref{compl-og-ax},
we have $W \vdash p' = q'$, so that the thesis $WB \vdash p = q$ follows by transitivity.
\fine
\end{theorem}

%
\section{Conclusion and Future Research}\label{conc-sec}
%

In this paper, we have shown how to use the process algebra SFM1, that truly represents NFAs up to isomorphism, 
in order to find a sound and complete axiomatization of 
language equivalence, and so also a sound and complete, finite axiomatization of regular languages, which is alternative
to those based on the Kleene algebra of regular expressions \cite{Salomaa,Conway,Kozen}. As a matter of fact,
our axiomatization, composed of 7 axioms and 3 conditional axioms, seems more concise than these.
On the one hand, the classic axiomatization 
by Kozen \cite{Kozen} is composed of 11 axioms for disjunction and concatenation, and 4 axioms for iteration 
(two of which are conditional). However, this is not the so far 
optimal axiomatization of language equivalence with Kleene algebras; in fact,
\cite{KZ20} showed that one of Kozen's conditional axioms can 
be removed at the price of adding a few equational axioms, and then \cite{DDP18} improved this result
by showing that language equivalence can be axiomatized by a (left-handed) Kleene algebra defined by 13 axioms, 
only one of which is conditional.
On the other hand, it should be remarked that our axiomatization makes use of axiom schemata (in particular,
for recursion), while \cite{Kozen,DDP18} do not.

NFAs without $\epsilon$-labeled transitions have been investigated by Alexandra Silva in her Ph.D. thesis \cite{Silva},
using co-algebraic techniques \cite{Rutten}. In chapter 5, based on a joint work with Bonsangue and Rutten \cite{SBR10},
she considers a (co-)algebra to describe such NFAs that, similarly to SFM1, features an action prefixing operator 
(but the prefix $\epsilon$ is not considered)
instead of concatenation, and recursion (implemented by means of the standard fixpoint operator of \cite{Mil84}) instead of Kleene star, but this algebra also 
contains a few additional operators. For this (co-)algebra, she proved a sort of representability theorem, up to language equivalence. 
Moreover,
she defines an axiomatization of language equivalence that, for the syntactic peculiarities of her (co-)algebra, turns out to be rather
verbose, as it is composed
of 18 axioms (while our axiomatization for SFM1 without the prefix $\epsilon$, is composed of 8 axioms only, 
as {\bf T3} and {\bf R3} are not necessary,
and, moreover, {\bf R2} can be replaced by the simpler {\bf R2$'$}).

NFAs without $\epsilon$-labeled transitions are also the model of another process algebra, called $BSP_{gfrec}(A)$ in \cite{BBR09} (cf. Corollary 5.8.3), 
an algebra rather similar to SFM1 (without the prefix $\epsilon$).
However, the axiomatic theory presented there is only for bisimulation equivalence and, instead of using axiom {\bf R2}, it resorts to some recursion principles,
in particular RSP ({\em recursive specification principle},  that ensures that each recursive specification has at most one solution), widely promoted in the ACP/BPA community.

Another goal of this paper is to show the usefulness
of representability theorems, up to isomorphism (like Theorem \ref{representability-nfa}), for the algebraic 
study of semantic models, like NFAs. In fact, SFM1 has allowed us to study also bisimulation equivalence on NFAs and to 
hint a sound and complete
axiomatization for it (the set $B$), composed of 4 axioms and 2 conditional axioms, only.
Future research may be devoted to extend the axiomatic theory of SFM1 to other behavioral equivalences in 
the linear-time/branching-time
spectrum \cite{vG01,GV15}. For instance, we claim that weak simulation equivalence \cite{Park81,AF+14} can be axiomatized by 
replacing axiom {\bf T2} in the set $W$ with the axiom 

{\bf T2$'$} \quad $a.(x+y) = a.(x+y) + a.y$.

Interestingly enough, we have proposed similar representability theorems
about various classes of Petri nets in \cite{Gor17},
offering for each class a specific process algebra truly representing the models of that class, up to isomorphism. 
In \cite{Gor19,Gor20ic} one of 
these results has been exploited for axiomatizing bisimulation-based, behavioral equivalences over a small 
class of finite Petri nets.

\begin{figure}[t]
\centering
    \begin{tikzpicture}[shorten >=1pt,node distance=2cm,on grid,auto]
    
       \node[accepting,state,label={below:$q_0$}]            (q0)               {};
      \node[accepting,state,label={below:$q_1$}] (q1) [right of=q0] {};

      \path[->] (q0) edge   [bend left]     node {$a$} (q1)
                (q1) edge  [bend left] node {$b$}        (q0);
       \node[accepting,state,label={above:$q_2$}] (q2)   [right of=q1]            {};
      \node[accepting,state, label={above:$q_3$}] (q3) [right of=q2] {};
      \node[state,label={above:$q_4$}]           (q4) [right of=q3] {};
      \node[state,label={left:$q_5$}]           (q5) [below of=q2] {};
        \node[accepting,state,label={right:$q_6$}]           (q6) [right of=q5] {};

      \path[->] (q2) edge node {$a$} (q3)
                    (q2) edge node {$a$} (q5)
                (q3) edge [bend left] node {$b$} (q4)
                  (q4)   edge  [bend left]   node {$a$} (q3)
                (q5) edge [bend left] node {$b$} (q6)
                (q6) edge [bend left] node {$a$} (q5);

    \end{tikzpicture}    
\caption{An NFA not representable in BPA$^*$, up to bisimilarity, in (a); and in (b) one expressible in BPA$^*$ that is language equivalent to the NFA in (a) }
\label{nfa-no-reg}
\end{figure}
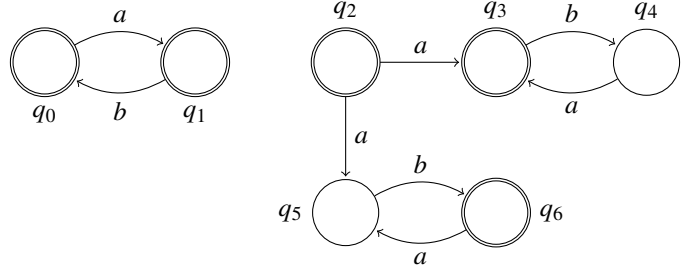

In the literature there are other process algebras that describe regular behaviors, the most prominent one being BPA$^*$ 
(see, e.g., \cite{BFP01,BCG07,GV15}), which corresponds syntactically to the definition of regular expressions. 
The abstract syntax of BPA$^*$ is
\[
p \; ::= \; \nil \mid \1 \mid \alpha \mid p + p  \mid p \cdot p \mid p^*
\]
where 
$p_1 \cdot p_2$ is the operator of {\em sequential composition},
and $p^*$ is the {\em iteration} operator. 
The semantic models used to give semantics to this process algebra are NFAs.
For instance, the semantics of $(a \cdot b)^* + a \cdot (b \cdot a)^*$
is isomorphic to the NFA outlined in Figure \ref{nfa-no-reg}(b). 
Note that the language recognized by
the NFA associated to the BPA$^*$ process $(a \cdot b)^* + a \cdot (b \cdot a)^*$ is exactly  
the language associated to the regular expression $(a \cdot b)^* + a \cdot (b \cdot a)^*$. 

Interestingly enough, this process algebra is strictly less expressive than SFM1.
In fact, Milner already observed in \cite{Mil84} that there are NFAs
that are not bisimulation equivalent to any process of BPA$^*$: for instance, the NFA in Figure \ref{nfa-no-reg}(a),
which instead can be represented in SFM1, up to isomorphism (hence, also up to bisimilarity), simply by the system of equations

$\begin{array}{rcl}
C_0 & \eqdef & a.C_1 + \1\\
C_1 & \eqdef & b.C_0 + \1\\
\end{array}$

However, BPA$^*$ can represent all the regular languages:
given a regular language $L$, there exists a BPA$^*$ process $p$ such that its associated NFA
recognizes $L$.  For instance,
a BPA$^*$ process, which is language equivalent to the one in Figure \ref{nfa-no-reg}(a), is actually
$(a \cdot b)^* + a \cdot (b \cdot a)^*$, whose associated semantics is isomorphic to the NFA in Figure \ref{nfa-no-reg}(b). 
This is not very surprising, as we already noted in the introduction (cf. Figure \ref{rep1}) that regular expressions are a formalism that can
represent all the NFAs, but only up to language equivalence. 
An interesting study about the subclass of NFAs that are actually expressible, up to bisimilarity,
with BPA$^*$ is described in \cite{BCG07}.

Our axiomatization is based on results and techniques developed in the area of process 
algebras \cite{Mil84,Mil89,HoPA,BBR09,San10,GV15} for bisimulation-based equivalences,
that we have applied and adapted to language equivalence, an equivalence relation
usually not deeply investigated in the context of process algebras.\\



\end{document}